\input harvmac

%
\figno=0
\def\fig#1#2#3{
\par\begingroup\parindent=0pt\leftskip=1cm\rightskip=1cm\parindent=0pt
\baselineskip=14pt \global\advance\figno by 1 \midinsert
\epsfxsize=#3 \centerline{\epsfbox{#2}} \vskip 12pt {\bf Fig.
\the\figno:} #1\par
\endinsert\endgroup\par
}
\def\figlabel#1{\xdef#1{\the\figno}}
\def\encadremath#1{\vbox{\hrule\hbox{\vrule\kern8pt\vbox{\kern8pt
\hbox{$\displaystyle #1$}\kern8pt} \kern8pt\vrule}\hrule}}

\overfullrule=0pt


\def\s{{\sigma}}
\def\o{\omega}

\def\a {{\alpha}}
\def\b {{\beta}}

\def\g {{\gamma}}

\def\d {{\delta}}
\def\e {{\epsilon}}

\def\Ind{\rm{Ind}}

\def\Todd{\rm{Todd}}
\def\Ch{\rm{Ch}}

\def\T{\rm T}

\def\t{\tau}
\def\a{\alpha}
\def\b{\beta}
\def\d{\delta}
\def\l{\lambda}
\def\th{\theta}
\def\g{\gamma}

\def\e{\epsilon}

\def\S{\Sigma}
\def\s{\sigma}

\def\be{\begin{eqnarray}}
\def\ee{\end{eqnarray}}

\Title{ {\vbox{ \rightline{}
 }}} {\vbox{\hbox{\centerline{The
moduli space of hyper-K{\"a}hler four-fold
compactifications}} }}

\bigskip \centerline{{\bf  Ram
Sriharsha}\footnote{$^{\dag}$}{ harsha@glue.umd.edu}}
\centerline{\it Department of Physics, University of Maryland}
\centerline{\it College Park, MD 20742-4111}

\bigskip
\bigskip
\baselineskip 18pt
\bigskip
\noindent

I discuss some aspects of the moduli space of hyper-K{\"a}hler
four-fold compactifications of type II and ${\cal M}$- theories. The
dimension of the moduli space of these theories is strictly bounded
from above. As an example I study Hilb$^2(K3)$ and the generalized
Kummer variety $K^2(T^4)$. In both cases RR-flux (or $G$-flux in
${\cal M}$-theory) must be turned on, and we show that they give
rise to vacua with ${\cal N}=2$ or ${\cal N}=3$ supersymmetry upon
turning on appropriate fluxes. An interesting subtlety involving the
symmetric product limit $S^2(K3)$ is pointed out.

\Date{December 1st, 2006}
\newsec{Introduction}

Compactifications of type II strings on hyper-K{\"a}hler two-folds
has been much studied and very well understood (see the review
\ref\AspinwallMN{
  P.~S.~Aspinwall,
  ``K3 surfaces and string duality,''
  arXiv:hep-th/9611137.} and references therein
for an excellent overview). The reason for this happy state of
affairs is that any two compact hyper-K{\"a}hler two-folds are
diffeomorphic to each other and there is essentially only one $K3$
surface. The moduli space of $K3$ surfaces can be determined
precisely, and in string theory we see this simplicity as the fact
that type IIA on a $K3$ surface is dual to heterotic strings on
$T^4$ whose moduli space is the Narain moduli space $Gr(4,20)$. An
analogous understanding of hyper-K{\"a}hler four-folds is lacking in
literature. In fact there are still only two known examples of
compact hyper-K{\"a}hler four-folds even though the cohomology of a
compact hyper-K{\"a}hler four-fold has been understood for a long
time. Any treatment of compactifications on hyper-K{\"a}hler
four-folds suffers from the fact that there are so few examples.
Fortunately, it turns out that one can map out the moduli space of
type II string theories on compact hyper-K{\"a}hler four-folds using
simple CFT arguments \ref\CecottiKZ{
  S.~Cecotti,
  ``N=2 Landau-Ginzburg versus Calabi-Yau sigma models: Nonperturbative
  Int.\ J.\ Mod.\ Phys.\ A {\bf 6}, 1749 (1991).}
. We review the argument that obtains the moduli space of ${\cal
N}=(4,4)$ SCFTs in Appendix 1 and use the results of {\CecottiKZ} to
obtain the moduli space of type IIA /B compactifications on
hyper-K{\"a}hler four-folds. There is an action of $O(4,b_2-2;{\bf
Z})$ on the moduli space of hyper-K{\"a}hler four-fold
compactifications of type IIA, which was observed by Verbitsky as
the group acting on the lattice $H^*(X;{\bf Z})$ for an arbitrary
hyper-K{\"a}hler manifold. We point out that there is a simple
reason why this group acts on $H^*(X;{\bf Z})$. Somewhat
surprisingly, it is possible to show that these theories have a
moduli space of bounded dimension, essentially due to the fact that
the topological types of hyper-K{\"a}hler four-folds is bounded. It
would be interesting to obtain a simple physical understanding of
this fact.

In section 2 the basic facts of hyper-K{\"a}hler four-folds is
summarized. Sections 3 and 4 work out aspects of the dimensional
reduction of type II theories on hyper-K{\"a}hler four-folds. In
most respects this is similar to the case of Calabi-Yau four-folds
and we follow the paper of Gates, Gukov and Witten \ref\GatesFJ{
  S.~J.~J.~Gates, S.~Gukov and E.~Witten,
  ``Two two-dimensional supergravity theories from Calabi-Yau four-folds,''
  Nucl.\ Phys.\ B {\bf 584}, 109 (2000)
  [arXiv:hep-th/0005120].}
in performing this reduction.

Some of the type IIA/${\cal M}$-theory compactifications will be not
be solutions of the 1-loop effective action coming from string
theory. We analyze this in more detail in section 5. In section 6 we
work out the conditions under which the two known examples of
hyper-K{\"a}hler four-folds yield supersymmetric vacua. We also show
that the symmetric product $S^2(K3)$ does not arise in the moduli
space of hyper-K{\"a}hler compactifications with fluxes.

Though we are only talking about hyper-K{\"a}hler four-folds in this
paper, it appears that there are few other ways of obtaining ${\cal
N}=3$ supersymmetric theories in 3d. In particular there is no other
class of compactifications that yield ${\cal N}=3$ supersymmetry in
3d at weak coupling. So one may view our results as indicating that
the moduli space of ${\cal N}=3$ theories in 3d (and the moduli
space of ${\cal N}=(3,3)$ supergravity theories in 2d) are very
tightly constrained. In particular the moduli space is of finite
dimension with a strict upper bound on the dimension. It has already
been noted that a similar thing happens for Calabi-Yau
compactifications in general, in that the dimension of the moduli
space of ${\cal N}=2$ supergravity theories in 4d for example is
expected to be finite. What is perhaps surprising is the simplicity
of showing this for ${\cal N}=3$ supergravities in 3d. In
particular, suppose we consider ${\cal N}=4$ supergravity in 3d, we
know that a class of these theories arise via compactification on
$K3\times K3$ whose moduli space is of finite dimension. However
these compactifications do not exhaust all ${\cal N}=4$ supergravity
theories in 3d and in particular a large class of such
compactifications arise via dimensional reduction on $T^2\times
CY_3$ for which there are only indirect arguments that suggest a
bound on the dimension of the moduli space.

Flux compactification in ${\cal M}$-theory and type II context has a
long history (see \ref\DouglasES{
  M.~R.~Douglas and S.~Kachru,
  ``Flux compactification,''
  arXiv:hep-th/0610102.} for overview and references)
. However, the recent work of Aspinwall and Kallosh
\ref\AspinwallAD{
  P.~S.~Aspinwall and R.~Kallosh,
  ``Fixing all moduli for M-theory on K3 x K3,''
  JHEP {\bf 0510}, 001 (2005)
  [arXiv:hep-th/0506014].}
on $K3\times K3$ is very closely related to the analysis in this
paper, and some of the techniques used there are applied to the case
of hyper-K{\"a}hler four-folds here.

\newsec{Some facts on Hyper-K{\"a}hler four-folds}

 A hyper-K{\"a}hler 4-fold is a K{\"a}hler manifold with a nowhere
vanishing non degenerate holomorphic 2-form ${\omega}$. Then
${\omega}^2$ trivializes the canonical line bundle, so by Yau's
proof of Calabi conjecture, there is a unique Ricci-flat metric that
respects the hyper-K{\"a}hler structure. The cohomology of a general
K{\"a}hler manifold can be decomposed via Hodge decomposition. For a
hyper-K{\"a}hler 4-fold, the non trivial Hodge numbers are
$h^{1,1},h^{2,1},h^{3,1}$ and $h^{2,2}$. However, not all of them
are independent. Given any type $(1,1)$-form we can create a $(3,1)$
form by wedging with ${\omega}$, so that $h^{3,1} = h^{1,1}$. Also
$h^{1,1} \geq 1$ as the space is K{\"a}hler, so we can write
$h^{1,1} =1 + p$ for some $p$ in ${\bf Z}^+$. Furthermore, just as
for a Calabi-Yau 4-fold, $h^{2,2}$ is not independent. The quickest
way to note this is to consider the index of the Dolbeault operator
${\bar {\partial}}_{E_2}$ acting on the bundle $E_2$  of holomorphic
type (2,0) forms. This index is given by:
\eqn\ai{ {\Ind}{\bar{\partial}}_{E_2} = {\sum}_{q=0} {(-1)}^q
h^{2,q} }

However, the index also has a purely topological character, and can
be expressed via the Atiyah-Singer Index theorem as:
\eqn\aii{ {\Ind}({\bar{\partial}}_{E_2}) = \int
{\Todd}(X){\Ch}({\Omega}^{2,0}) }

Using the standard expression for the Todd genus and Chern
character, we compute:
\eqn\aii{ {\Ind}({\bar {\partial}}_{E_2}) = {1\over {120}} \int
(3{c_2}^2 + 79c_4 ) }

where we used the fact that $c_1 = 0$. Now, the Todd genus of a
hyper-K{\"a}hler 4-fold is precisely 3, and this implies a relation
between ${c_2}^2$ and $c_4$ (incidentally, $\int c_4$ is the
arithmetic genus or Euler characteristic of the hyper-K{\"a}hler
4-fold $X$). Specifically:
\eqn\aiii{ {\Todd}(X) = {1\over {720}}(3{c_2}^2 - c_4) }

so that
\eqn\aiv{ \int_X {c_2}^2 = 720 + {{\chi}\over 3}}

Using {\aiv} in {\aii} we get a relation between the various Hodge
numbers. Denoting $h^{2,1} = 2q$ \footnote{$^1$}{Here we used the
fact that $b_3$ is divisible by 4, for a hyper-K{\"a}hler four-fold.
Incidentally, this also implies ${\chi}$ is divisible by 12, which
is a stronger result than the one for Calabi-Yau four-folds. The
Hilbert scheme of two points on K3 gives us an example where
${\chi}$ is divisible by 12, and not by 24, so this is the strongest
result we can get. In our notation ${{\chi}\over {24}} = {1\over
2}(7 + p -q)$.} this relation is:
\eqn\av{ h^{2,2} = 72 + 8p - 4q}

So the hyper-K{\"a}hler 4-folds are characterized by two non
negative integers $(p,q)$.

\newsec{ Compactification of type IIA on Hyper-K{\"a}hler four-folds}

In this section we will describe the compactification of type IIA
string theory on a hyper-K{\"a}hler four-fold $X$. In the large
volume limit these compactifications can be discussed by
dimensionally reducing type IIA supergravity on hyper-K{\"a}hler
four-folds.

The bosonic content of type IIA supergravity in ten dimensions is
the metric $g_{MN}$, an antisymmetric two-form $B_{MN}$ and dilaton
${\phi}$ from the NS-NS sector. The R-R sector gives rise to the
one-form gauge field $A_M$ and three form $C_{MNP}$. The bosonic
action in string frame is of the form:
\eqn\avi{L = \int d^{10}x {\sqrt{-g}}[e^{-2{\phi}}(R^{10}+
4({\nabla}{\phi})^2 - {1\over {12}}H^2) - {1\over 4}F^2 - {1\over
{48}}G^2] }

Where:
\eqn\avii{F =dA \quad H = dB \quad G = dC + A\wedge H}
are the gauge invariant field strengths. The action {\avi} is of
course the tree level action for type IIA string theory in ten
dimensions. There are higher order terms in the effective action
that are not captured in {\avi}. For the most part their structure
is not known. There is however an important term of the form
$B\wedge X_8$ where $X_8$ is a particular contraction of four powers
of the Riemann tensor. This term was shown to be present in type IIA
by considering scattering amplitudes in type II string theory
\ref\vafa{C.~Vafa and E.~Witten,
  ``A One Loop Test Of String Duality,''
  Nucl.\ Phys.\ B {\bf 447}, 261 (1995)
  [arXiv:hep-th/9505053].}
. This term leads to a tadpole for the $B$-field which has to be
cancelled in type IIA by turning on $G$-flux and/ or adding $N$
F1-strings such that:
\eqn\aviii{ N = {{\chi}\over {24}} - {1\over {2(2{\pi})^2}}\int_X
G\wedge G}

If the Euler number of $X$ is not divisible by 24, then the tadpole
cannot be canceled by simply adding F-strings and we must turn on
RR-flux $G$ also. Of course, turning on $G$-flux we will typically
end up breaking supersymmetry unless the $G$-flux happens to be
primitive with respect to the ${\bf P}^1$ of complex structures on
$X$. For the moment we will ignore these subtleties and address them
in section 3. The action for type IIA supergravity is invariant with
respect to 32 supercharges, 16 of which are left-chiral and 16
right-chiral with respect to the chirality operator in 10d. Upon
compactifying on $X$, the resulting action in two dimensions
possesses residual supersymmetry only if $X$ admits a covariantly
constant spinor. In the case of hyper-K{\"a}hler four-folds the
holonomy group of $X$ is sp(2). A generic eight dimensional spinor
is in one of the two inequivalent spinor representations of spin(8)
say ${\bf 8}_+$. Under sp(2) we have the decomposition:
\eqn\aix{ {\bf 8}_+ = {\bf 5} + {\bf 1} + {\bf 1} + {\bf 1} \quad
{\bf 8}_- = {\bf 4} + {\bf 4}}

so that there is a three-dimensional space of covariantly constant
spinors on $X$. Via the decomposition:
\eqn\ax{{\bf 16} = (8_+,+) + (8_-,-), \quad {\bf 16}' = (8_+,-) +
(8_-,+), }

corresponding to $SO(1,9) \rightarrow SO(8) \times SO(1,1)$ we end
up with a non-chiral two dimensional supergravity
theory\footnote{$^2$}{ In {\ax} the ${\bf 16}$ and ${\bf 16}'$ refer
to the ten dimensional Majorana-Weyl spinors of opposite chirality
associated to type IIA, whereas the spinor representations of
$SO(1,1)$ are labeled by their charges under $spin(1,1)$.} with
${\cal N}=(3,3)$ supersymmetry upon compactifying type IIA on $X$.

To determine the spectrum of the resulting two dimensional theory
one performs Kaluza-Klein reduction of the various fields of type
IIA. As the resulting two dimensional theory is non-chiral the
fermions simply arise as ${\cal N}=(3,3)$ superpartners and it is
enough to count the massless bosonic degrees. These are associated
to the harmonics of the various bosonic fields of type IIA. Denoting
the holomorphic 2-form on $X$ by ${\omega}$, one can expand the
$B_{MN}$ zero modes as:
\eqn\axi{B = \sum_i b^i {\omega}^{1,1}_i + b{\omega} }

where:
\eqn\axii{{\omega}^{1,1} \in H^{1,1}(X) \quad b \in {\bf C} \quad
b^i \in {\bf R} }
leading to $h^{1,1} + 2$ scalars. The $C_{MNP}$ zero modes lead to
$2h^{2,1}$ scalars and $h^{1,1} + 2$ vectors via:
\eqn\axiii{C = \sum_j c^j{\omega}^{2,1}_j + \sum_n
{C_{\mu}}^n{\omega}^{1,1}_n + C_{\mu}{\omega} \quad
{\omega}^{2,1}\in H^{2,1}(X) \quad c^j \in {\bf C} }

The metric deformations lead to $3h^{1,1} -2$ scalars $g^k$ as
follows: The zero modes of the graviton satisfy the Lichnerowicz
equation which in a suitable gauge can be written as:
\eqn\axiv{D_kD^k h_{ij} - R_{isjt}h^{st} = 0}
It is easy to see that the metric variations of the form
${\d}h_{ab}$ and ${\d}h_{a{\bar b}}$ do not mix in {\axiv} so they
can be considered separately. For every element ${\omega}^{1,1}$ one
obtains a variation of the form ${\d}h_{a{\bar b}}$ so that the
number of such deformations is $h^{1,1}$. Similarly, given
${\omega}^{1,1} \in H^{1,1}(X)$ one can construct a variation of
type ${\d}h_{ab}$ as:
\eqn\axv{{\d}h_{ab} = {{\omega}^{\bar c}}_{(a}
{\omega}^{1,1}_{b){\bar c}} }

However if ${\omega}^{1,1}$ is proportional to the K{\"a}hler form
then {\axv} vanishes, so that there are only $2h^{1,1}-2$
deformations of type ${\d}h_{ab}$ so that the space of sp(2)
holonomy metrics on a hyper-K{\"a}hler four-fold has dimension
$3h^{1,1} -2$.

Collecting all the matter content together we end up with $h^{1,1} =
(p+1)$ ${\cal N}=(4,4)$ vector multiplets containing $g^k,b_i$ as
the scalar components, together with $q$ ${\cal N}=(4,4)$ hyper
multiplets containing the $4q$ scalars $c^j$. Even though we have
only ${\cal N}=(3,3)$ supersymmetry, the matter sector arranges
itself into ${\cal N}=(4,4)$ multiplets, which is a familiar fact
given that any supersymmetric sigma model with ${\cal N}=3$
supersymmetry is automatically ${\cal N}=4$ supersymmetric also. Of
course the higher order terms in the effective action will only be
${\cal N}=(3,3)$ supersymmetric.

The supergravity sector contains the graviton, three abelian gauge
fields and a scalar, along with three gravitini and three Majorana
fermions. The dilaton sits in the supergravity multiplet.

The low energy effective action for the vector and hyper-multiplet
moduli will in general be given by a ${\cal N}=(4,4)$ supersymmetric
sigma model. In the case of the vector multiplets with rigid
supersymmetry this sigma model is based on a target space that is
hyper-K{\"a}hler with torsion (HKT), so we expect upon coupling to
supergravity that the target space is quaternionic K{\"a}hler with
torsion (QKT). The hyper multiplet moduli space is similarly a
hyper-K{\"a}hler or Quaternionic K{\"a}hler manifold. As the two
multiplets carry scalars with different R-symmetries the moduli
space factorizes just as in ${\cal N}=2$ supergravity coupled to
matter in four dimensions. Denoting the moduli space ${\cal M}$ of
type IIA on a hyper-K{\"a}hler four-fold as:
\eqn\axvii{{\cal M} = {\cal M}_V \times {\cal M}_H}
what can be said about ${\cal M}_V$ and ${\cal M}_H$?

The worldsheet description of any ${\cal N}=(3,3)$ supersymmetric
compactification to two dimensions is in the form of a ${\cal N}=4$
SCFT with small ${\cal N}=4$ SCA and $c=12$. The space-time moduli
that sit in the $(p+1)$ vector multiplets are all ${\cal N}=4$
chiral primary operators of this internal ${\cal N}=4$ SCA. Since
any ${\cal N}=4$ SCA has a $SU(2)_L\times SU(2)_R$ R-symmetry this
implies \footnote{$^3$}{ Details of this standard argument are
provided in appendix 1. This argument was first applied for
determining the moduli space of ${\cal N}=4$ SCFTs by Cecotti.} that
the moduli space ${\cal M}_V$ has a $SO(4)$ isometry. It turns out
due to a theorem of Berger and Simons (see {\AspinwallMN} for a nice
discussion on the Berger-Simons result) that the smooth manifolds
with $SO(4)$ holonomy and dimension greater than 4 are only the
symmetric spaces, the so called Grassmann manifolds. This leads us
to identify:
\eqn\bi{{\cal M}_V = {{O(4,p+1)} \over {{O(4)\times O(p+1)}}}}

There is a natural $O(4,p+1;{\bf Z})$ symmetry of the moduli space
which we can quotient by maintaining the Hausdorff property of the
the resulting space. It is natural to conjecture that the U-duality
group for this theory is $O(4,p+1;{\bf Z})$.

In type IIA the dilaton ${\phi}$ is in the ${\cal N}=(3,3)$
supergravity multiplet. This implies that the form of the moduli
space is completely independent of string coupling $g_s = e^{\phi}$.
For large $g_s$, type IIA goes over to 11d supergravity which is the
low energy limit of ${\cal M}$-theory. This means the ${\cal
M}$-theory moduli space is also given by {\bi}. By the same
argument, the metric on the moduli space is independent of string
coupling.

Given the moduli space of the form {\bi}, we can take the large
radius limit. The large radius limit can be determined by examining
the Dynkin diagram of $O(4,p+1)$, and it turns out that the
structure of the moduli space in the large radius limit is given by:
\eqn\biii{ {\cal M} = {{O(3,p)} \over {{O(3)\times O(p)}}} \times
{\bf R}_+ \times {\bf R}^{p+3}}

This is what we expect in the large radius limit. In this limit we
expect the metric deformations to be characterized by the moduli
space of $sp(2)$ holonomy metrics of fixed volume of a
hyper-K{\"a}hler four-fold, which is the $O(3,p)$ factor, the ${\bf
R}_+$ factor corresponds to the trivial radial mode. The ${\bf
R}^{p+3}$ factor corresponds to the scalars arising from dimensional
reduction of the NSNS 2-form. This provides a non-trivial
consistency check.

The ${\cal N}=(3,3)$ supergravity coupled to matter has not been
constructed in literature. There is however the case of ${\cal
N}=(4,4)$ supergravity coupled to matter which has been analysed
\ref\PerniciDQ{
  M.~Pernici and P.~van Nieuwenhuizen,
  ``A Covariant Action For The SU(2) Spinning String As A Hyperkahler Or
  Phys.\ Lett.\ B {\bf 169}, 381 (1986).}. This theory has a gauged $SU(2) \in SO(4)$ R-symmetry and it has
been shown that the target space parameterized by the scalars in
this theory can be hyper-K{\"a}hler or Quaternionic K{\"a}hler. We
expect a similar result to hold even in the case of ${\cal N}=(3,3)$
supergravity coupled to matter. That is, with ${\cal N}=(3,3)$
supersymmetry, the form of the moduli space remains non-trivial in
general. This raises the puzzle as to how the CFT analysis was able
to determine the local form of the moduli space as {\bi}. We will
resolve this puzzle in the next section.

One subtlety that has to be pointed out is that there is a
difference between the $K3$ case and the case of general
hyper-K{\"a}hler manifolds which affects our understanding of the
moduli space. For $K3$ surfaces the global Torelli theorem holds, so
that the moduli space of complex structures is determined by the
space of periods. It is the space of periods that the supergravity
analysis is sensitive to, and so is the chiral primary ring of the
${\cal N}=(4,4)$ worldsheet theory. It is not known whether a
version of the global Torelli theorem holds for the higher
dimensional cases. If it does not, then the choice of periods does
not determine the complex structure fully. What will be lacking is
some discrete data. It is known that all hyper-K{\"a}hler manifolds
are  deformations of a projective variety so they all have ${\pi}_1
= 0$. So it is not possible to have discrete torsion \ref\VafaWX{
  C.~Vafa,
  ``Modular Invariance And Discrete Torsion On Orbifolds,''
  Nucl.\ Phys.\ B {\bf 273}, 592 (1986).}
 in the
worldsheet SCFT. I do not know what extra data the SCFT can have in
this case that is not captured by the chiral primary ring. So the
analysis of the  moduli space in this paper is carried out modulo
the discrete ambiguity arising from lack of a global Torelli like
theorem.

\subsec{ ${\cal M}$-theory on hyper-K{\"a}hler four-folds}

The low energy limit of ${\cal M}$ theory is 11d supergravity whose
bosonic content is a graviton and a 3-form potential $A$, with
four-form flux $G$.

Dimensional reduction of 11d supergravity on a hyper-K{\"a}hler
four-fold yields a three dimensional ${\cal N}=3$ supergravity
coupled to matter. The matter multiplets are the vector multiplet (
whose bosonic content is  three scalars transforming as ${\bf 3}$ of
the SO(3) R-symmetry together a gauge field) and the hyper multiplet
(which contains four scalars transforming as a complex doublet of
the R-symmetry). Any action for the hyper multiplets is
automatically ${\cal N}=4$ supersymmetric, so is the low energy
effective action for the vector multiplets (in the absence of
G-flux). Upon dimensional reduction, we end up with a ${\cal N}=3$
supergravity multiplet with a graviton, three gravitini. The matter
sector consists $p+1$ vector multiplets (after dualizing some
vectors into scalars) and $q$ hyper multiplets. The moduli space
factorizes as in the type IIA case. Upon dualizing the vectors into
scalars, we expect the ${\cal M}$-theory moduli space to coincide
with the type IIA case. The ${\cal M}$-theory moduli space will be
of the form:
\eqn\biv{{\cal M}_{11d} = {{O(4,p+1)}\over {O(4)\times O(p+1)}}
\otimes
  {\cal M}_H}

\newsec{ Compactification of type IIB on hyper-K{\"a}hler four-folds}

The compactification of type IIB string theory on a hyper-K{\"a}hler
four-fold $X$ leads to a two dimensional ${\cal N}=(0,6)$
supersymmetri theory in the non-compact directions. Its low energy
limit is ${\cal N}=(0,6)$ supergravity coupled to matter. In this
section we determine the matter content of this theory and the
moduli space. In the large volume liimit type IIB string theory in
ten dimensions is well approximated by type IIB supergravity. The
bosonic content of type IIB supergravity is the graviton $g_{MN}$,
the anti-symmetric two form $B_{MN}$, the dilaton ${\phi}$, the RR
axion $C$, along with the RR two form A$_{MN}$ and the self-dual
four-form $G_{MNPQ}$. Type IIB in ten dimensions has a $sl_2({\bf
Z})$ action where the two forms $A,B$ form a doublet of $sl_2({\bf
Z})$ and the axio-dilaton can be combined as:
\eqn\hi{ {\t} = c + ie^{-\phi} }
and transforms under $sl_2({\bf Z})$ as:
\eqn\hii{ {\t} \rightarrow {{(a{\t} + b)}\over {(c{\t} + d)}}\quad
a,b,c,d \in {\bf Z} \quad ad-bc = 1}

As the five form field strength:
\eqn\hiii{ F = dG + {3\over 4}B \wedge dB }
is self dual, there is no covariant action whose equation of motion
yields the self-duality constraint. Agreeing to impose this
constraint by hand, we can write down a lagrangian for type IIB
supergravity. As in the type IIA case, we need to determine the
massless spectrum of particles in the 2d theory. The NS-NS sector
modes $g,B$ and ${\phi}$ give rise to the same zero modes for both
type IIA and IIB. So we end up with $4h^{1,1} + 1$ scalars from the
NS-NS sector.

In type IIA we argued that the dilaton went into the supergravity
multiplet. In type IIB it is the fluctuation of the radial mode of
the metric that goes into the ${\cal N}=(0,6)$ supergravity
multiplet.

From the RR sector, the axion gives rise to a real scalar. The RR
two form $A$ gives rise to $h^{1,1} + 2$ scalars exactly as the
B-field. The expansion of the self-dual five-form $F$ is more
complicated. It can be expanded as follows:
\eqn\hiv{ F = \sum dC^i {{{\omega}^4}_-}_i  + \sum d{C'}^j
{{{\omega}^4}_+}_j }
where ${\omega}^4_-$ refers to the space of anti self-dual
four-forms on $X$, while ${\omega}^4_+$ refers to the space of
self-dual four-forms. The self-duality of $F$ implies that $C^i$ are
anti self-dual scalars, while $C'^j$ are self-dual. This means the
scalars $C^i$ are left-moving while the scalars $C'^j$ are right
moving. In this notation the supersymmetries of the 2d theory are
purely right-moving.

Therefore, in the purely right-moving matter sector we have $b^4_- +
5h^{1,1} + 3$ scalars.

The middle dimensional cohomology of $X$ decomposes into self-dual
and anti self-dual pieces by Poincare duality. The signature ${\s}$
of $X$ is nothing but ${\s} = b^4_+ - b^4_-$. The Hirzebruch
signature theorem relates ${\s}$ to the Euler character of $X$:
\eqn\hv{ {\s} = {1\over {45}}\int (7p_2 - p_1^2) = 48 + {{\chi}\over
3} }
Furthermore, the Euler formula gives:
\eqn\hvi{2(b^0 + b^2 + b^3) + b^4_+ + b^4_- = {\chi}  }

Using {\hv} and {\hvi} together with the relation ${\chi} =
12(7+p-q)$ we can easily determine:
\eqn\hvii{b^4_- = 3(h^{1,1}-1) = 3p }

That is, we end up with $n =8h^{1,1} = 8(p+1)$ right moving scalars
which by ${\cal N}=(0,6)$ supersymmetry have $8(p+1)$ right moving
Majorana fermions as superpartners. Again this is consistent with
the fact that in the rigid supersymmetry limit the dimension of the
target space of the right-moving moduli must be a multiple of 8.

Of course, to complete the spectrum we need to compute the left
moving fields as well, but since they will play no part in the rest
of the discussion we will not explicitly count the left-movers.
Suffice it to say that they ensure that the resulting two
dimensional theory is free from gravitational anomalies.

The ${\cal N}=(0,6)$ supergravity coupled to $8n$ matter multiplets
has not been constructed in literature.  The important point about
this theory is that the target space for the scalars is completely
fixed, even though the theory has only six supercharges, it behaves
more like the case of ${\cal N}= 4$ supergravity with 16
supercharges in four dimensions. There is a simple argument to see
why the target space for the right-moving moduli is fixed by $n$, in
the case of ${\cal N}=(0,6)$ supergravity . It starts out with the
observation that in the case of rigid supersymmetry, any sigma model
with ${\cal N}=(0,6)$ supersymmetry is based on a flat target space
(upto orbifolding by a discrete group). The reason for this is
simple: with ${\cal N} = (0,6)$ supersymmetry and beyond, the only
super multiplet possbile with this much supersymmetry has scalars
transforming non trivially under the R-symmetry that rotates the
supercharges. In the ${\cal N}=(0,8)$ case for example the scalars
form a ${\bf 8}_v$ of the spin(8) R-symmetry, whereas in the ${\cal
N}=(0,6)$ case the scalars form a ${\bf 4}$ of $SU(4)\sim SO(6)$.
Every such sigma model if it were non trivial would give rise to a
conformally invariant theory in the IR, with ${\cal N}=6$ SCA and
above. However there is no superconformal extension of the ${\cal
N}=6$ supersymmetry algebra. This means the IR theory must be scale
invariant without being conformally invariant and therefore every
such sigma model should actually correspond to a free
theory\footnote {$^4$}{In the non compact case one can have scale
invariance without conformal invariance essentially by turning on
the dilaton, but this is not possible in the compact case.}.

Indeed in the ${\cal N}=(0,6)$ case the scalars transform in the
${\bf 4}$ of $SU(4)$ so there are actually $8n$ scalar fields rather
than a multiple of 6 which would have required the scalars to
transform in the fundamental of $SO(6)$. As we have argued above, in
the case of rigid ${\cal N}=(0,6)$ supersymmetry the target space
parameterized by the right-moving scalars is actually flat and is
simply ${\bf R}^{8n}$ locally (in our case $n = p+1$). This means
that any non-trivial moduli space arises for these scalars precisely
by coupling to ${\cal N}=(0,6)$ supergravity. Upon coupling to
${\cal N}=(0,6)$ supergravity there is a mass parameter ${\kappa}$
that essentially plays the role of the gravitational Newton's
constant in 4d (the 2d gravitational coupling is dimensionless).
When ${\kappa} \rightarrow 0$ the target space becomes flat ${\bf
R}^{8n}$ and for non-zero ${\kappa}$ the target space for the
right-moving moduli must have a curvature proportional to
${\kappa}$. All of this is analogous to what happens for ${\cal
N}=2$ supergravity coupled to matter in four dimensions. In this
case in the rigid supersymmetry limit the target space for the
scalars must be hyper-K{\"a}hler, whereas local supersymmetry
requires the target space to be quaternionic K{\"a}hler with
negative curvature proportional to the 4d Newton's constant. The
only difference is that for ${\cal N}=(0,6)$ supergravity in 2d, the
rigid supersymmetry limit is trivial and this we expect will put
severe constraints on the moduli space arising out of local
supersymmetry. In particular, this moduli space can be exactly
determined. The actual construction of ${\cal N}=(0,6)$ supergravity
coupled to matter will be explored in a forthcoming paper.

This allows us to resolve the puzzle raised at the end of the
previous section. Even though supergravity does not drastically
constrain the moduli space of type IIA compactifications to three
dimensions (with six supercharges), it turns out that supergravity
does constrain the moduli space of type IIB compactification. Since
type IIA and IIB are related to each other upon compactifying one
dimension further (and T-dualizing) this provides us with an
understanding of why the type II moduli space for hyper-K{\"a}hler
four-fold compactifications can be determined locally.

We claim that the type IIB moduli space ${\cal M}$ is given by:
\eqn\biii{{\cal M} = {\T}^*{{O(4,p+1)} \over {{O(4)\times O(p+1)}}}}

That is locally ${\cal M}$ is a bundle over the Grassmannian with
fibers ${\bf R}^{4p+4}$. Globally of course, the fibers are compact,
and are actually tori (which is what makes the moduli space compact,
assuming the T-duality group acts on the base). The form of the
moduli space we expect from the large radius limit is:

\eqn\biv{{\cal M} = {\bf R}^{5p+6} \times {{SL_2({\bf R})}\over
{U(1)}} \times {{O(3,p)} \over {{O(3)\times O(p)}}}}

We have been schematic in writing {\biv} and it should be thought of
as a warped product of the individual factors. Since the radial mode
goes into the ${\cal N}=(0,6)$ supergravity multiplet, we expect the
large radius limit to be exact.

The moduli space at weak coupling (the CFT moduli space) is of the
form:
\eqn\bv{{\cal M} = {\bf R}^{4p+4} \times {{O(4,p+1)} \over
{{O(4)\times O(p+1)}}}}

which agrees with the topology of {\biii}. One way to show {\biii}
is to simply dimensionally reduce type IIB on hyper-K{\"a}hler
four-folds in a manner similar to  and observe that the moduli space
is of the form of a cotangent bundle over the space parameterized by
the metric and B-field deformations. The cotangent bundle structure
follows exactly as in the analysis of Gates, Gukov and Witten . This
structure of course arises rather straightforwardly upon dimensional
reduction, but our claim is that this form of the moduli space is
fixed by the ${\cal N}=(0,6)$ supergravity of the 2d theory.

\subsec{ T-duality}

The space {\biii} has a base which can be thought of as the space of
space-like four planes in ${\bf R}^{4,p+1}$. The group $O(4,p+1)$
naturally acts on ${\bf R}^{4,p+1}$ into which we can embed an
integral lattice ${\Lambda}^{4,p+1}$. The subgroup $O(3,p)$ of
$O(4,p+1)$ is the rotation of the integral lattice
${\Lambda}^{3,b_2-3}$ of $H^2$. In type IIB there is also the
$sl_2({\bf Z})$ duality inherent in ten dimensions. This leaves the
lattice ${\Lambda}^{3,b_2-3}$ untouched but mixes the base and
fibers. It was pointed out by Verbitsky \ref\verb{M.~Verbitsky,
``Cohomology of compact hyperk{\"a}hler manifolds,''
[alg-geom/9501001].}that there is a group action on the integral
cohomology lattice of any hyper-K{\"a}hler $2n$-fold of the form
$SO(4,b_2-2)$ which in particular holds for four-folds. The
$O(4,p+1)$ factor can thus be identified with the symmetry of the
integral cohomology lattice of the four-fold. This motivates the
$O(4,p+1;{\bf Z})$ duality group of type IIA. In type IIB this has
to be extended by the action of $sl_2({\bf Z})$.

In the type II theories D-brane charges are vectors in the lattice
$H^*(X;{\bf Z})$ and the action of $SO(4,b_2-2)$ that acts as an
automorphism of this lattice rotates the D-brane charges the way
T-duality is supposed to work. This leads us to suspect that
$SO(4,p+1;{\bf Z})$ is nothing but the T-duality group of the type
IIA theory.

Indeed $SO(4,p+1;{\bf Z})$ is the T-duality group of the worldsheet
SCFT corresponding to hyper-K{\"a}hler four folds. However as we
will see soon some of these compactifications are destabilized by
the 1-loop correction. In those cases the T-duality group may be
strictly smaller. The fact that the classical T-duality group is
$SO(4,p+1;{\bf Z})$ ties in neatly with the observation that
$SO(4,p+1)$ acts on $H^*(X)$ via the fact that D-brane charges are
Mukai vectors in the lattice $H^*(X;{\bf Z})$.

\subsec{ Dimension of the ${\cal M}$-theory moduli space}

The dimension of the moduli space of type IIA(or ${\cal M}$-theory)
compactification on hyper-K{\"a}hler four-folds is set by two
integers $p$ and $q$. However it is easy to show that there is an
upper bound on $p$ for any hyper-K{\"a}hler four-fold.

In fact, Beauville has shown that $b_2 \leq 23$ which implies $p
\leq 20$. This restriction follows from the observation that $
Sym^2(H^2) \hookrightarrow H^4 $.

From this we note:
\eqn\bvi{ b_2(1+b_2) \leq 2b_4 }

Furthermore, by an Index theorem of Salamon (we have written it for
hyper-K{\"a}hler four-folds, though the Index theorem holds for all
hyper-K{\"a}hler manifolds):
\eqn\bvii{ b_4 = -b_3 + 10b_2 + 46 }

Using {\bvi} and {\bvii} we get:
\eqn\bviii{ (b_2 - 23)(b_2 + 4) \leq 0}

which implies $b_2 \leq 23$. In the case where $b_2 = 23$ the
inclusion map $i:Sym^{2i}(H^2) \hookrightarrow H^{4i}$ is exact and
gives the only non vanishing Hodge numbers leading to the Hodge
diamond of the Hilbert scheme.

Rather non-trivially even the integer $q$ is bounded from above for
hyper-K{\"a}hler four-folds by a number that depends on $p$
\ref\Guan{D.~Guan, ``On the betti numbers of Irreducible compact
hyperK{\"a}hler manifolds of complex dimension four,''Math. Research
Letters 4(2001),663-669.}. Indeed the analysis of {\Guan} concludes
that not every topological type of hyper-K{\"a}hler four-fold is
possible. Either $b_2 =23$ in which case the hyper-K{\"a}hler
four-fold has the same Hodge diamond as Hilb$^2(K3)$, or $b_2 \leq
8$. Furthermore, for all $b_2 \leq 8$, $b_3$ is bounded above by a
number that depends on $b_2$.

It is clear that {\bvi} and {\bvii} also put a bound on $q$ but the
bounds derived in {\Guan} are much stronger.

Including the moduli coming from the position of membranes, the
entire moduli space is of bounded dimension. A similar situation
arises for ${\cal N}=4$ supersymmetric string compactifications in
four dimensions, but it is nice to see the moduli space of theories
with six supercharges is bounded by a calculable finite number. It
would be very interesting to understand a more physical reason for
the precise bound on $p$ and $q$. Of course, this would follow from
the ${\cal N}=4$ worldsheet SCFT by analyzing the Elliptic genus,
together with the CFT version of {\bvi}, but perhaps there is a more
compelling reason, just as the dimension of the ${\cal N}=4$
theories in 4d were restricted by the rank of the gauge group that
could come out of the Heterotic string.

\newsec{Effect of Fluxes}

In the previous section we considered hyper-K{\"a}hler
compactifications where we set all RR fluxes to zero, and indeed
only the metric degree of freedom was excited. It turns out that in
general this does not lead to consistent string propagation. In
fact, there is a 1-loop correction to the B-field equation of motion
of type IIA, which in general requires the RR 4-form flux to be
turned on to solve. This is simplest to understand in the ${\cal
M}$-theory context where Witten \ref\WittenMD{
  E.~Witten,
  ``On flux quantization in M-theory and the effective action,''
  J.\ Geom.\ Phys.\  {\bf 22}, 1 (1997)
  [arXiv:hep-th/9609122].}
 observed that the $G$-flux of
${\cal M}$-theory does not obey Dirac quantization condition. If we
consider the periods of the $G$-flux on a spin four-fold, then
Witten showed:
\eqn\ci{ [{G\over {2\pi}}] - {1\over 2}{\l} \in {\bf Z} }

where ${\l} = {1\over 2}p_1$. In other words $G$ obeys usual
quantization condition precisely when ${\l}$ is even. If however
${\l}$ is odd, then it is clear from {\ci} that $G$ can not be taken
to vanish. We will see that for the two known hyper-K{\"a}hler
four-folds, this condition is not satisfied, so they do not solve
the string equations of motion until one turns on $G$-flux. The
condition {\ci} is closely related to the tadpole condition in
${\cal M}$ theory \ref\VafaTF{
  C.~Vafa and E.~Witten,
  ``A Strong coupling test of S duality,''
  Nucl.\ Phys.\ B {\bf 431}, 3 (1994)
  [arXiv:hep-th/9408074].}
, where the Bianchi identity of $G$ is corrected as:
\eqn\cii{ d*G = {1\over {4{\pi}^2}}G\wedge G  - {{({p_1}^2 -
4p_2)}\over {192}}}

In fact, {\cii} upon integrating over the internal manifold has to
be integral (so that it can be canceled by adding membranes).
However, it can be checked {\WittenMD} that the class $[{G\over
{2\pi}}]$ is integral precisely when ${\l}$ is even according to
{\cii}, confirming {\ci}. For a complex four-fold with $c_1 = 0$:
\eqn\cii{ (p_1^2 - 4p_2) = 8{\chi}, }

so that {\ci} implies $G = 0$ is consistent only if ${\chi}$ is
divisible by 24. As we already noted, for a hyper-K{\"a}hler
four-fold:
\eqn\ciii{ {{\chi}\over {24}} = {1\over 2}(7 + p -q) }

Equation {\cii} is a tadpole for the gauge field $C$ which couples
to the membrane charge in ${\cal M}$-theory. The charge conservation
relation following from {\cii} is:
\eqn\ciia{ n + {1\over {2(2\pi)^2}}\int_X G\wedge G = {{\chi}\over
{24}}}

As a result of {\ciia} we need to include membranes or turn on
$G$-flux to cancel the tadpole for $C$. The effect of turning on
fluxes will be discussed later in this section, and we will now
simply consider the effect of adding $n$ membranes spanning the
noncompact directions. There will now be additional moduli arising
from the fact that the $n$ membranes can be placed at arbitrary
points. These moduli are trivial to count, and including them the
${\cal M}$-theory moduli space becomes a warped product of the form:
\eqn\ciiaa{ {\cal M}_{11d} ={{O(4,p+1)}\over {O(4)\times O(p+1)}}
\otimes Sym^n(X)}

Therefore the effect of adding membranes is to increase the moduli.
However the effect of turning on fluxes is quite the opposite.

In the large radius limit the conditions for preserving
supersymmetry upon turning on $G$-flux was analyzed by Becker and
Becker \ref\BeckerGJ{
  K.~Becker and M.~Becker,
  ``M-Theory on Eight-Manifolds,''
  Nucl.\ Phys.\ B {\bf 477}, 155 (1996)
  [arXiv:hep-th/9605053].}
.

The result is that the $G$-flux must be primitive and of Hodge type
$(2,2)$. The primitivity condition can be written as:
\eqn\ciiab{ G\wedge J = 0 }
where $J$ is the K{\"a}hler form. This of course preserves only four
supercharges, as the freedom to rotate the ${\bf P}^1$ of complex
structures is broken by $G$. In order to preserve the full ${\cal
N}=3$ supersymmetry in the ${\cal M}$-theory compactification
{\ciiab} must hold for all the complex structures on the four-fold
$X$.

The condition for the form to be of type $(2,2)$ follows as usual
from (upon imposing self-duality of $G$):
\eqn\ciiac{G \wedge {\Omega} = 0}

where ${\Omega}$ is the holomorphic 4-form. As the holomorphic four
form ${\Omega}$ is simply the square of the holomorphic two form
${\omega}$, the condition {\ciiac} is equivalent to:
\eqn\ciiad{G \wedge {\omega} = 0}

In general the space of primitive type (2,2) forms is not easy to
describe. Given a generic HK 4-fold there may not be a non-trivial
space of such forms. However for any HK 4-fold we can show the
existence of at least one such form $G_ 0 = {\o}\wedge {\bar {\o}} -
{1\over 2}J^2$. Due to the explicit appearance of ${\o}$ in $G_0$
such a $G$-flux breaks ${\cal N}=3$ supersymmetry down to ${\cal
N}=2$.

The primitivity of $G_0$ follows from the fact that $g_{{\bar
a}b}{\o}_{bc} = {\bar {\o}}_{{\bar a}c}$. This can be seen as
follows.

Any hyper-K{\"a}hler manifold $M$ has a triplet of complex
structures $J^A$.

Every such complex structure is a anti-symmetric rank-2 tensor map
$J:TM \rightarrow TM$ which squares to $-1$, that is $J^i_j J^j_k =
-{\d}^i_k $. Furthermore, the almost complex structure defined by
$J$ is integrable, so that extend the definition of $J$ from the
tangent space to a point on $M$ to the entire manifold $M$. On a
hyper-K{\"a}hler manifold the triplet of complex structures $J$ are
required to satisfy the Lie algebra of $sp(1)$. That is:
\eqn\ciiadi{ J^A J^B = {\e}^{ABC} J_C }

Once a complex structure is picked, we can write $J_1 = J$ and $J_2
+ iJ_3 = {\o}$ and $J_2 - iJ_3 = {\bar {\o}}$ and {\ciiadi} becomes:
\eqn\ciiadii{ J^i_j {\o}^{jk} = {\bar {\o}}^{ik} }

where $(i,j,k)$ indices refer to the real coordinates on $M$.
Passing to complex coordinates $(a,b,c)$ we note that ${\o}$ is of
type (2,0) with respect to $J$ and $J$ is of type (1,1) and in
appropriate coordinates it can be expressed as $J^a_b = ig^a_b$
where $g_{a{\bar b}}$ is the K{\"a}hler metric. This means:
\eqn\ciiadiii{ J^{{\bar a}b}{\o}_{bc} = {\bar{\o}}^{\bar a}_c }

Now consider $i_JG_0$. Using:
\eqn\ciiadiv{ J_{{\bar a}b}{\o}^{bc} {\bar {\o}}^{{\bar a}{\bar c}}
= g^{c{\bar d}}{\bar \o}_{{\bar a}{\bar d}} {\bar \o}^{{\bar a}{\bar
c}} }
 and applying {\ciiadi} we infer:
 \eqn\ciiadv{ i_J {\o}\wedge {\bar {\o}} = J }

 That is:
 \eqn\ciiadvi{ i_J ({\o}\wedge {\bar {\o}} - {1\over 2}J^2) = 0 }

 Equation {\ciiadvi} implies $i_JG_0 = 0$. For a middle dimensional form like
 $G_0$, $i_J G_0 = 0$ implies $J \wedge G_0 = 0$. That is $G_0$ is
 primitive as we claimed. Note that ${\o}\wedge {\bar {\o}}$ is
 linearly independent from $J^2$ for hyper-K{\"a}hler manifolds with
 quaternionic
 dimension greater than 1, so $G_0$ is not vacuous. For a $K3$
 surface of course ${\o}\wedge {\bar{\o}}$ is a linear multiple of
 $J\wedge J$ as the space of 4-forms on a $K3$ surface is one
 dimensional.

It appears that the choice of $J$ in $G_0$ fixes the K{\"a}hler
structure of the manifold. As we will see, the choice of ${\o}$
fixes the complex structure up to an overall scale ${\l} \in {\bf
C}^* $ and the $G$-flux can be taken to be of the form ${\mu}G_0$ so
that ${\o} \rightarrow {\l}{\o}$, ${\mu}\rightarrow
|{\l}|^{-2}{\mu}$ leaves $G_0$ fixed if $J \rightarrow |{\l}| J$ so
that there is still a residual ambiguity corresponding to the radial
modulus which is not fixed by the choice of this flux.

If however $G$ satisfies both {\ciiab} and {\ciiad} and is not of
the form $G_0$ then it is automatically primitive with respect to
the entire ${\bf P}^1$ of complex structures.

Turning on $G$ flux of the form $G_0$ constrains the complex
structure. In fact the complex structure is entirely fixed by the
choice of $G$-flux in this case.

Given ${\o}$ we write ${\o} = {\a} + i{\b}$ where $({\a},{\b}) \in
H^2(X;{\bf R})$. For a $K3$ surface the middle dimensional
cohomology was the lattice ${\bf Z}^{22}$ which was even, unimodular
and had a quadratic form with signature (3,19) associated to it. For
a higher dimensional hyper-K{\"a}hler manifold most of this
structure does not generalize. However, for any hyper-K{\"a}hler
manifold $X$ we can look at the cohomology $H^2(X;{\bf Z})$. There
is a quadratic form $q_X$ called the Beauville-Bogomolov
form\footnote{$^5$}{ An excellent introduction to compact
hyper-K{\"a}hler manifolds is the lectures of Huybrechts \ref\Huy{
D.~Huybrechts, ``Compact hyperk{\"a}hler manifolds: basic results,''
preprint 1997.} and the paper of Beauville \ref\Beau{A.~Beauville,
``Varietes K{\"a}hleriennes dont la primiere classe de Chern est
nulle,'' J.\ Diff.\ Geom {\bf 18}, 755 (1983). } where the quadratic
form is introduced. Integrality follows by a result of Fujiki
\ref\Fuj{A.~Fujiki, ``On the de Rham cohomology of a compact
K{\"a}hler symplectic manifold,'' Adv.\ Stud.\ Pure Math. {\bf 10}
105, (1987).}.} associated to $H^2(X;{\bf Z})$ making it into a
lattice ${\Lambda}^{3,b_2-3}$ which is integral but not necessarily
even or unimodular.

In the case of Hilb$^2(K3)$ the lattice ${\Lambda}^{3,20} = E_8
\oplus E_8 \oplus H^3 \oplus (-2{\bf Z})e $ and with a suitable
normalization the Beauville-Bogomolov quadratic form is even and
integral.

The Hodge-Riemann identities are:
\eqn\ciiae{ q_X({\o},{\o}) = 0 \quad q_X({\o},{\bar {\o}}) > 0 }

In terms of ${\a}$ and ${\b}$ we have:
\eqn\ciiaf{ q_X({\a},{\a}) = q_X({\b},{\b}) , \quad q_X({\a},{\b}) =
0 }

That is, the choice of a holomorphic 2-form ${\o}$ is equivalent to
the choice of a space-like 2-plane spanned by the periods of ${\a}$
and ${\b}$ in the period domain ${\Lambda}^{3,b_2-3}$. Together with
$J$ this determines a space-like 3-plane ${\cal O}$ in
${\Lambda}^{3,b_2-3}$.

The choice of flux $G_0$ is automatically a choice of ${\o}$ and $J$
so it corresponds to a choice of the space-like 3-plane ${\cal O}$
inside ${\Lambda}^{4,b_2-2}$. By supersymmetry the periods of the
$B$-field is also fixed by this choice of $G$-flux. However
splitting off ${\Lambda}^{4,b_2-2} = {\Lambda}^{3,b_2-3} \oplus
{\Lambda}^{1,1}$ we see that fixing the space-like 3-plane ${\cal
O}$ still leaves unfixed the radial mode( in fact one complex
dimension  if we include the ${\cal N}=2$ superpartner)remains
unfixed. It is still non-trivial to find a $G_0$ that satisfies the
flux quantization condition. We will examine this in detail in the
next section in the context of an example.

\subsec{ Extra constraints on moduli}

In general, there are only three equations arising from the
supergravity constraints {\ciiab} and {\ciiac}. When the $G$-flux is
not of type $G_0$ it appears therefore that only a triplet of moduli
can be removed at each instance, whereas with ${\cal N}=3$
supersymmetry we expect quaternionic dimensions to disappear. This
means there is more moduli being removed than governed by {\ciiab}
and {\ciiac}. In the ${\cal M}$-theory setting this happens because
certain modes of the 3-form $C$ are constrained due to the
Chern-Simons coupling $C\wedge G\wedge G$.

Indeed upon turning on $G$-flux this Chern-Simons coupling leads in
a standard fashion to the $3d$ Chern-Simons action for the zero mode
of $C$ so that the $G$-flux appears to give topological mass to the
vector field $C_{\mu}$ sitting in one of the $p$ vector multiplets.
Together with the mass terms for the triplet of scalars, this is
enough to lift precisely one quaternionic dimension.

The 11d supergravity action is of the form:
\eqn\ciifi{ S = \int d^{11}x {\sqrt{-g}}(R -{1\over 2}F\wedge *F) -
{1\over {24{\pi}^2}}C\wedge F\wedge F  }

Expanding the 3-form in harmonics:
\eqn\ciifii{ C = {\sum}_a {\o}_a A^a_{\mu} + {\sum}_i {\o}^{2,1}_i
A^i  \quad {\o}_a \in H^2(X;{\bf R}) }

we end up with a 3d Chern-Simons action:
\eqn\ciifiii{ S = {1\over {4{\pi}}} \int d^3x {\l}_{ab}A^a \wedge
F^b , \quad {\l}_{ab} = \int_X {\o}_a \wedge {\o}_b \wedge {G\over
{\pi}}  }

In the case where ${\l}$ is not even, ${G\over {\pi}}$ is an
integral class, so {\ciifiii} as normalized is ${1\over 2}$ of the
canonical Chern-Simons action in 3d. Furthermore, ${\l}_{ab}$ as
defined in {\ciifiii} is integral. The Chern-Simons action on a
3-manifold $W$ is defined by computing the Maxwell action:
\eqn\ciifiv{ S = {1\over {{2\pi}}} \int_Z{F\wedge F} }
for an arbitrary closed 4-manifold $Z$ with boundary $W$ by choosing
an extension of the gauge field on $Z$. The action {\ciifiv} is
independent of the choice of $Z$ modulo ${2\pi}$. Suppose the
4-manifold $Z$ is spin, then if $L$ is a complex line bundle over
$Z$ with $c_1(L) = {F\over {2\pi}}$ then $c_1^2(L)$ is divisible by
2 by Wu's formula (as the second Steiffel-Whitney class $w_2$
vanishes). That is, given a 3-manifold with a chosen spin structure,
the Chern-Simons action:
\eqn\ciifv{ S = {1\over {4\pi}}\int A\wedge F }
is the basic action (the so called level-${1\over 2}$ Chern-Simons
action). This agrees with the normalization in {\ciifiii}.

Going back to {\ciifiii}, we notice that the effect of turning on
background $G$-flux is to give topological mass to the gauge fields.
By ${\cal N}=3$ supersymmetry {\ciifiii} is related to mass terms
for the ${\cal N}=3$ superpartners. The ${\cal N}=3$ vector
multiplet in 3d consists of a vector and three scalars. In ${\cal
N}=2$ notation we write the ${\cal N}=2$ vector multiplet as ${\S}$
and the chiral multiplet as ${\Phi}$. ${\S}$ contains a real scalar
and a vector which together with the chiral multiplet form the
content of a ${\cal N}=3$ vector multiplet (the theory is parity
invariant under ${\l}_{ab} \rightarrow -{\l}_{ab}$ so the multiplet
is the same as a ${\cal N}=4$ vector multiplet). In terms of this
the superpotential can be scehmatically written as:
\eqn\ciifvi{ S = \int d^3x d^4{\th} {\l}_{ab}{\S}^aV^b -\int
d^3xd^2{\th}i{\l}_{ab}{\Phi}^a{\Phi}^b , \quad {\S} = iD{\bar D}V}

{\ciifvi} is only schematic since we have ignored the coupling to
gravity and as written {\ciifvi} is simply the ${\cal N}=3$
supersymmetric Chern-Simons action.

For non-zero $G$, parity invariance in 3d is broken by the
Chern-Simons coupling {\ciifiii}. This is simply because the
$G$-flux is odd under 11d parity, and any expectation value breaks
parity in 11d, and upon compactification in the resulting 3d theory
also.

The analysis leading to {\ciifiii} is really independent of the
details of the internal manifold which are subsumed in ${\l}_{ab}$.
Suppose we consider $K3\times K3$. In this case the 3d theory has
${\cal N}=4$ supersymmetry which prevents the appearance of a
Chern-Simons term. However, as shown in {\AspinwallAD} it is
possible to turn on $G$-flux consistent with ${\cal N}=4$
supersymmetry. Indeed as we saw above, the 11d Chern-Simons coupling
gives rise to a topological mass for the gauge fields irrespective
of the precise amount of supersymmetry, so we should expect this
coupling to be present even for $K3\times K3$. However, it is well
known that there is no ${\cal N}=4$ supersymmetric Chern-Simons
action. There is in fact only one way to complete {\ciifiii} in a
manner consistent with ${\cal N}=4$ supersymmetry. To explain this
let us consider the dimensional reduction on $K3\times K3$.

Upon dimensionally reducing {\ciifi}  the gauge fields $A^a_{\mu}$
arise via reduction of $C$ on $K3\times K3$. An equal number of such
gauge fields arise via dimensional reduction on either $K3$. Instead
of considering all those gauge fields together as $2(h^{1,1} + 2)$
vector multiplets of the ${\cal N}=4$ supersymmetry, we can rather
consider them as $h^{1,1}+2$ vector multiplets and $h^{1,1}+2$
twisted vector-multiplets. Doing so the 11d Chern-Simons coupling
leads upon dimensional reduction to a BF type coupling between the
vector and twisted vector-multiplets, lifting a pair of quaternionic
dimensions at a time.

To be more precise, the ${\cal N}=4$ supersymmetry algebra in three
dimensions has a $SU(2)_R\times SU(2)_N$ R-symmetry, the eight super
charges being doublets under the two R-symmetry factors. The ${\cal
N}=4$ vector multiplet has a vector, three real scalars transforming
as ${\bf 3}$ of $SU(2)_R$ as bosonic components. The ${\cal N}=4$
supersymmetry algebra admits an automorphism that exchanges the two
$SU(2)$ factors and takes a vector multiplet into a so-called
twisted vector-multiplet which has three scalars that transform as
${\bf 3}$ of $SU(2)_N$. In the ${\cal N}=4$ supergravity that arises
upon compactifying ${\cal M}$-theory on $K3\times K3$, the two
$SU(2)$ factors can be related to the holonomies of the $K3$s. In
fact, upon compactifying ${\cal M}$-theory on a product of
four-manifolds $Y\times Y$, the holonomy group is $SO(4)\times
SO(4)$ generically, leading to the absence of R-symmetries in the
resulting 3d theory (which is not supersymmetric unless $Y$ has
reduced holonomy). Suppose $Y$ is a $K3$ surface, then decomposing
$SO(4)$ as $SO(4) = SU(2)\times SU(2)$ the holonomy of $Y$ can be
taken to be one of the two $SU(2)$ factors, and the other $SU(2)$
factor therefore becomes an R-symmetry. The same thing happens with
the other factor of $Y$ thus leading to a $SU(2)_R\times SU(2)_N$
R-symmetry as noted. The important point is that the two $SU(2)$
factors are associated with the two $K3$ surfaces.

With this identification, it is clear that the $2(h^{1,1}+2)$ vector
multiplets that arise by dimensional reduction of the 3-form have to
be treated as $(h^{1,1}+2)$ vector multiplets and $(h^{1,1}+2)$
twisted vector-multiplets as claimed, because the scalars in these
multiplets transform under different $R$-symmetries. There is a
unique renormalizable coupling that involves vector and twisted
vector-multiplets and is called the BF coupling \ref\KapustinHA{
  A.~Kapustin and M.~J.~Strassler,
  ``On mirror symmetry in three dimensional Abelian gauge theories,''
  JHEP {\bf 9904}, 021 (1999)
  [arXiv:hep-th/9902033].}
. It is precisely this coupling that arises via dimensional
reduction of ${\cal M}$-theory to 3d.

Again schematically the ${\cal N}=4$ superpotential can be written
as:
\eqn\ciifvii{ S = \int d^3xd^4{\th}{\l}_{aa'}{\S}^a{\tilde V}^{a'} -
i{\l}_{aa'}{\Phi}^a{\tilde{\Phi}}^{a'} }

where ${\S},{\Phi}$ form a ${\cal N}=4$ vector multiplet and
${\tilde{\S}}$ and ${\tilde{\Phi}}$ form a ${\cal N}=4$
twisted-vector multiplet.

\newsec{Examples}

Examples of compact hyper-K{\"a}hler four-folds are very hard to
obtain. There are only two known examples in literature, and both of
them are obtained from symmetric products of complex 2-folds (a K3
surface in the case of the Hilbert scheme, and $T^4$ in the example
of Beauville).

\subsec{The Hilbert scheme of two points on K3  }

A K3 surface is a compact (simply connected) K{\"a}hler surface with
trivial canonical bundle (its holonomy is $SU(2) = sp(1)$ and is in
fact a hyper-K{\"a}hler manifold of complex dimension 2). As all
$K3$ surfaces are diffeomorphic, one can compute the unique Hodge
numbers of a K3 surface by picking a suitable representative. The
Fermat form of the quartic in ${\bf CP}^3$ is a simple example given
by:
\eqn\civ{ S: z_1^2 + z_2^2 + z_3^2 + z_4^2 = 0 \quad z_i \in {\bf C}
\quad z_i \sim {\l} z_i \quad {\l} \in {\bf C}^*}

By the adjunction formula one can compute the Chern classes of the
surface $S$ using its embedding in ${\bf CP}^3$:
\eqn\cv{ c(T_S)c(N_s) = c({T{\bf CP}^3}_{|S}) }

Using the fact that
\eqn\cvi{ c({T{\bf CP}^3}_{|S}) = (1 + x)^4 \quad c(N_S) = (1 + 4x)
\quad x \in H^2({\bf CP}^3,{\bf Z}) }

We compute $c_1(S) = 0$ and $c_2(S) = 6x^2$. The Lefschetz
hyperplane theorem tells us $S$ is simply connected, so $S$ is
actually a K3 surface as advertised. Its Euler characteristic is:
\eqn\cvii{ {\chi} = \int_S c_2 = 24\int_{{\bf CP}^3} x^3 = 24}

The Euler characteristic also has an expression in terms of Hodge
numbers as:
\eqn\cviii{{\chi} = h^{1,1} + 2h^{2,0} + 2 }

allowing us to read off $h^{1,1} = 20$, since we know $h^{2,0} = 1$.

Given a K3 surface $X$, there is a construction by
Beauville\footnote{$^6$}{ A nice introduction to hyper-K{\"a}hler
manifolds can be found in Beauville's lectures
\ref\beauville{A.~Beauville, ``Riemannian Holonomy and Algebraic
Geometry,'' [arXiv:math/9902110].}}, that allows us to obtain a
compact hyper-K{\"a}hler four-fold. Starting from a K3 surface $X$,
the symmetrized product $S^2(X)$ admits a symplectic structure that
is derived from the holomorphic 2-form present on $X$.

However, as it stands there are other symplectic structures also,
and we need to be able to construct a unique (upto scaling)
symplectic structure to obtain a hyper-K{\"a}hler 4-fold. This can
be done by considering $X^2 = S^2(X)/G_2$ where $G_2$ is the
symmetric group of order 2. This space is singular, but has a nice
de singularization into what is called a Hilbert scheme $X^{[2]}$
which parameterizes finite subspaces of $X$ of length $2$. The
Hilbert scheme can be thought of as a resolution of the singular
space $X^2$, and the holomorphic 2-form survives this resolution and
yields a smooth, compact hyper-K{\"a}hler four fold $X^{[2]}$.

One can determine $b_2(X^{[2]}) = b_2(X) + 1$ and $b_3(X^{[2]}) =
0$, so that the Hodge numbers of $X^{[2]}$ can be determined as
follows:
\eqn\cix{ h^{1,1} = 21 \quad h^{2,1} = 0 \quad h^{2,2} = 232 }

This allows us to conclude that ${{\chi}\over {24}} = {{27}\over
2}$, so we need to turn on $G$-flux to satisfy the $G$-equation of
motion in ${\cal M}$-theory. This implies a destabilization of the
purely gravitational background. We need to turn on $G$-flux, and
furthermore if we want to preserve any supersymmetry the $G$-flux
has to be primitive.

As we have mentioned before, the form ${G\over \pi} = {\mu}G_0$ is
primitive. Inserting the form of this flux into {\ciia} we obtain:
\eqn\cx{ n + {1\over 8}{\mu}^2 \int_X (G_0/{\pi})^2 = {{27}\over 2}
}

We also require:
\eqn\cxa{{{G}\over{\pi}} \in H^4(X;{\bf Z})}

Only if {\cx} and {\cxa} simultaneously have solutions do we have a
consistent string background at this order. To preserve
supersymmetry $n$ has to be non-negative.

Let us analyze this more carefully.

First of all, let us fix the normalization of $G_0$ to make it
primitive. It is clear that $i_J ({\o}\wedge {\bar{\o}})$ is
proportional to $J^2$ so we need to fix the constant of
proportionality. For this we can use the fact that:
\eqn\qi{ {\int}_X J^4 = 3q^2_X(J,J) \quad {\int}_X {\o}{\bar{\o}}J^2
= q_X({\o},{\bar{\o})q_X(J,J) }}

so that:
\eqn\qii{{G\over {\pi}} = {\mu}({\o}\wedge{\bar {\o}} -
{{q_X({\o},{\bar{\o}})}\over {3q_X(J,J)}} J\wedge J )}

is primitive and type (2,2). Now it is easy to compute{\cx} to be:
\eqn\qiii{ n + {5\over {24}}{\mu}^2q^2_X({\o},{\bar {\o}}) =
{{27}\over 2}}

which has a solution with $n=6$ and $q_X({\o},{\bar {\o}}) = 2$ and
${\mu} = {\pm}3$.

Flux quantization required ${\o}\wedge {\bar{\o}}$ to be integral
(we simply absorb ${\mu}$ into ${\o}$ by scaling ${\o}\rightarrow
{\l}{\o}$ with ${\l} = {\sqrt{\mu}}$).

As mentioned before, the HK moduli space is lifted save for the
complex dimension in which the radial mode resides. There are
however moduli arising from the position of the membranes in the
transverse eight dimensional space.

The above analysis raises the following puzzle: in the orbifold
limit of $K3^{[2]}$ studied by Dasgupta, Rajesh and Sethi
\ref\DasguptaSS{
  K.~Dasgupta, G.~Rajesh and S.~Sethi,
  ``M theory, orientifolds and G-flux,''
  JHEP {\bf 9908}, 023 (1999)
  [arXiv:hep-th/9908088].}, it was found that there was no way of turning
 on $G$-flux in a supersymmetric manner for the orbifold ${K3}^2$
whereas we have just argued that ${K3}^{[2]}$ does have a
supersymmetric solution upon turning on suitable $G$-flux.

Let us try to locate the precise nature of this discrepancy. In
{\DasguptaSS} the orbifold limit of the symmetric product of $K3$
was considered. In the limit where $K3$ could be described by
$T^4/{\bf Z}_2$ as:
\eqn\cxaai{ g_1: (z^1,z^2) \rightarrow -(z^1,z^2) }

They obtained the condition for a primitive type $(2,2)$ $G$-flux
to solve the anomaly cancelation condition as:
\eqn\cxaaii{ 2({|A|^2 + B^2 + C^2}) + n = {{27}\over 2} }

where the primitive type (2,2) flux was written down as :
\eqn\cxaaiii{ {G\over {2\pi}} = Ad{\bar z}^1dz^2d{\bar z^3}dz^4 +
Bd{\bar z}^1dz^2dz^3d{\bar z}^4 + Cd{\bar z}^1d{\bar z}^2dz^3dz^4 +
h.c }

The condition that the period of ${G\over {2\pi}}$ be half integral
quantized gave the restriction:
\eqn\cxaaiv{ (ReA \pm B {\pm} C ) \in {\bf Z}  \quad ImA \in {\bf Z}
}

Due to the symmetrization $(B,C)$ are real. Now it was noted in
{\DasguptaSS} that there is no solution to both {\cxaaii} and
{\cxaaiv}. This means the orbifold limit of $K3^{[2]}$ is not a
supersymmetric solution. This seems to directly contradict our claim
that $K3^{[2]}$ preserves supersymmetry as a type IIA or ${\cal
M}$-theory compactification. Indeed there is something puzzling
about the fact that {\cxaaii} and {\cxaaiv} do not agree.

To be precise, any solution to the flux quantization condition
allows us to solve {\cxaaiv} for an integral number of branes. In
the orbifold limit $S^2({\bf T}^4/{\bf Z}_2)$ the flux quantization
condition simply requires the periods of ${G\over {2\pi}}$ to be
integral (or half-integral). Away from the quotient singularities
the space is locally flat, and ${\l}$ can be taken to vanish.

In the absence of torsion, this would imply the quantization
condition {\cxaaiv}. Now in the case of compact eight-manifolds, it
was shown by Witten that any $G$ flux solving the flux quantization
condition also solves the anomaly cancelation condition with
integral number of branes. This statement does not carry over to the
orbifold as we just saw above. This is somewhat surprising.

As we saw above, we can turn on $G$-flux with ${\cal N}=2$
supersymmetry preserving vacua. In this case we lift all the complex
structure moduli, and we're left with one K{\"a}hler moduli and the
associated  period of the B-field. That is there is a ${\bf R}$
worth of B-field periods. Now $S^2(K3)$ with $G$-flux turned on has
atleast one B-field period that is unfixed, this corresponds to the
volume modulus of the $K3$ itself. We will see this to imply that
the orbifold point $S^2(K3)$ does not exist in the ${\cal N}=2$
moduli space.

To be more precise, we have to formulate what it means to reach the
symmetric product point $S^2(K3)$. The moduli space of complex
structures on Hilb$^2(K3)$ has dimension 20, which is one greater
than the moduli space of complex structures of $K3$. This means at a
generic point in moduli space the internal manifold is not of the
form $K3^{[2]}$. Geometrically as we mentioned before, $K3^{[2]}$ is
obtained from $S^2(K3)$ by blowing up the exceptional divisor $e \in
H^{1,1}(K3^{[2]};{\bf R})$. In fact $q_X(e,e) = -2$ so $e$ is a
time-like vector in ${\Lambda}^{3,20}$. A choice of complex
structure is the same as a choice of a positive 3-plane in
${\Lambda}^{3,20}$ and this induces a polarization of $e$ as $e =
e^{3,0}_+ + e^{0,20}_-$ where the ${\pm}$ serve to indicate the
projection into space-like and time-like parts. To reach the point
in moduli space where we have $S^2(K3)$ we need to ensure that $e$
is orthogonal to the 3-plane spanned by $(J,{\o})$. In other words,
one has to rotate the 3-plane spanned by $(J,{\o})$ such that it is
orthogonal to $e$. However with ${\cal N}=2$ supersymmetry
preserving $G$-flux turned on, the complex structure is entirely
frozen. Indeed the 3-plane spanned by $(J,{\o})$ is fixed by the
choice of $G_0$ so it is no longer possible to reach the symmetric
product point.

Let us determine solutions with ${\cal N}=3$ supersymmetry. Let us
first choose $G ={\nu}{\a}^2$ where ${\a}$ is an element of
$H^{1,1}(X;{\bf Z})$.

This can be shown as follows: Pick primitive (2,2) form $G$ as:
\eqn\cxxd{ {G\over {\pi}} = {\nu} {\a}\wedge{\a} = {\nu} {\a}^2
\quad {\nu} \in 2{\bf Z}+1 }

where ${\a} \in H^2(X;{\bf Z}) \bigcap H^{1,1}(X;{\bf R})$ is
primitive. This means that ${\a}$ is orthogonal to the 3-plane
spanned by $({\o},J)$ in the lattice ${\Lambda}^{3,b_2-3}$ and is
furthermore time-like.

The anomaly cancelation condition now becomes:
\eqn\cxxe{ n + {{{\nu}^2}\over 8}\int_X{\a}^4 = {{27}\over 2} }

We can re-write {\cxxe} as:
\eqn\cxxf{ n + 3{{\nu}^2\over 8}q_X^2({\a},{\a}) = {{27}\over 2} }

As $q_X$ is even, we can write $q_X({\a},{\a}) = 2k$ so that we are
searching for solutions to:
\eqn\cxxfa { n + {3\over 2} {\nu}^2k^2 = {{27}\over 2} }

Clearly there is a non vanishing space of solutions to {\cxxfa}. One
such solution without any membranes has $k = 1$ and ${\nu} =
{\pm}3$.

So far we have not checked whether ${\a}^2$ is primitive. That is,
we require $i_J{\a}^2 = 0$. This actually cannot happen for our
choice of $G$ for any ${\a}$.

In general we can write:
\eqn\cui{ i_J{\a}^2 = {\g} }

Further, we can consider:
\eqn\cuii{{G\over {\pi}} = {\a}^2 - J\wedge {\g} - {1\over
2}aJ\wedge J }

where we denote by $a$ the contraction ${\a}^{i{\bar j}}{\a}_{i{\bar
j}}$.

It is easy to see that $i_JG = 0$ if $G$ is defined as in {\cuii}.
This allows $G$ to be a primitive type (2,2) form. However there is
an explicit dependence on $J$ so the choice of this form does not
leave us with the freedom to rotate the space-like 3-plane ${\cal
O}$, which means this choice also breaks ${\cal N}=3$ supersymmetry.

Let us therefore consider a $G$-flux of the form ${G\over {\pi}} =
{\a}\wedge {\b}$ with $({\a},{\b})$ in $H^{1,1}(X;{\bf Z})$ and
orthogonal to ${\cal O}$. Such $({\a},{\b})$ lie in the Picard
lattice of $X$. Under what condition is $G$ primitive? Clearly when
${\a} = {\b}$ such a $G$-flux cannot be primitive as we saw above.
However, supposing ${\a}$ and ${\b}$ are linearly independent, is it
possible to arrange for the $G$-flux defined above to be primitive?

In fact, consider ${\a}\wedge{\b}\wedge J = A$. Now suppose $x \in
H^2(X;{\bf R})$ be an arbitrary form, then we can show that $\int_X
A\wedge x = 0$ which automatically implies $A = 0$ that is $G$ is
primitive if and only if ${\a}$ and ${\b}$ are orthogonal with
respect to $q_X$. That is, a $G$-flux of the form:
\eqn\cuiii{ {G\over{\pi}} = {\a}\wedge {\b} , \quad {\a},{\b} \in
H^{1,1}(X;{\bf R})\bigcap H^2(X;{\bf Z}), \quad q_X({\a},{\b}) =
q_X({\a},J) = q_X({\b},J) = 0 }

is primitive and of type (2,2) and turning on such a flux preserves
${\cal N}=3$ supersymmetry.

That is, pick ${\a}$ and ${\b}$ in ${\Lambda}^{3,20}$, mutually
orthogonal, and orthogonal to ${\cal O}$. Given such ${\a}$ and
${\b}$ the $G$-flux defined in {\cuiii} is primitive and type (2,2)
and preserves ${\cal N}=3$ supersymmetry as the choice of such a
$G$-flux does not affect the rotations in ${\cal O}$.

In order to solve the anomaly cancelation condition we require:
\eqn\cuiv{ n + {1\over 8}q_X({a},{\a})q_X({\b},{\b}) = {{27}\over 2}
}

There are many solutions to {\cuiv} once we note that
${\Lambda}^{3,20}$ is an even lattice so {\cuiv} always has
solutions.

For Hilb$^2(K3)$ the space $H^4$ is comprised entirely of
$sym^2(H^2)$ so there are no other possibilities for $G$-flux that
give rise to ${\cal N}=3$ supersymmetry.

Turning on a $G$-flux of the form {\cuiii} reduces the dimension of
the moduli space. In fact, the moduli space is:
\eqn\cuv{ {\cal M}_V = {{O(4,19)}\over{O(4)\times O(19)}} }

We can find more possibilities by considering more general $G$-flux
as a linear combination of pairs of the form {\cuiii} and each time
we reduce the dimension of the moduli space by two quaternionic
dimensions at a time.

The fact that there exist compactifications with ${\cal N}=3$
supersymmetry forces us to reconsider the puzzle posed earlier. In
this case, it seems possible to reach the symmetric product point
while preserving ${\cal N}=3$ supersymmetry, which seems to
contradict the results of {\DasguptaSS}. A possible resolution to
this may be the following: the symmetric product treated using usual
orbifold techniques requires a certain ${\th}$ angle to be turned
on. In the limit discussed in {\DasguptaSS}, this was implicit. If
suppose the ${\th}$ angle is fixed at zero upon turning on $G$-flux,
then we cannot reach the limit discussed in {\DasguptaSS}.

\subsec{singularities}

It is an interesting problem to classify the type of singularities
that occur in the moduli space of hyper-K{\"a}hler four-fold
compactifications. In the case of $K3$ surfaces it is well known
that the only singularities that can occur in the $K3$ moduli space
are orbifold singularities which have an ADE classification. An
analogous understanding of higher dimensional hyper-K{\"a}hler
manifolds is lacking. Following its definition in the $K3$ case we
can define the Picard lattice of a HK 4-fold $X$ as:
\eqn\si{ Pic X = H^2(X;{\bf Z}) \bigcap H^{1,1}(X;{\bf Z}) }

Suppose we have an element ${\a}$ of the lattice ${\Lambda}^{3,20}$
of $K3^{[2]}$ which is orthogonal to ${\cal O}$, the 3-plane spanned
by $(J,{\o})$. Then ${\a}$ lies in $Pic(K3^{[2]})$. A generic
$K3^{[2]}$ has a vanishing Picard group, so the existence of such
${\a}$ is a restriction on the complex structure of $K3^{[2]}$.
Every element ${\a}$ of the Picard lattice gives rise to a line
bundle $L$ over $K3^{[2]}$ with $c_1(L) = {\a}$. Indeed the Picard
group is nothing but the group of such line bundles with the product
given by Whitney product formula. Any such line bundle is Poincare
dual to a divisor $L$. The zero section of such a line bundle will
describe a divisor which we also call $L$. The existence of a zero
section of the line bundle $L$ can sometimes be figured out using
the Riemann-Roch formula\ref\Gott{G.~Ellingsrud,L.~Gottsche,M.~Lehn,
``On the cobordism class of the Hilbert scheme of a surface,''
[arXiv:math.AG/9904095].}:
\eqn\sii{ {\chi}(L) = {1\over 2}({1\over 2}q_X({\a},{\a})+3)({1\over
2}q_X({\a},{\a}) + 2)  }

For line bundles over surfaces ${\chi} = h^0(L) - h^1(L) + h^2(L)$
and if ${\chi} > 0$ this implies $L$ (or $L^{-1}$) has a section.
This is no longer the case for four-folds. It would be interesting
to understand when a section exists. If it does, the corresponding
divisor will have zero volume. The volume of such a divisor L is
$\int_L J^3$ which is nothing but $\int_X J^3\wedge {\a}$. However,
we can easily show that $\int_X J^3\wedge {\a} = q_X({\a},J)q_X(J)$
upto an irrelevant numerical constant. Now $q_X({\a},J) =0$ so this
implies volume of $L$ vanishes. Such choices of ${\a}$ will then
lead to singular HK 4-folds.

It is well known that $Pic(K3^{[2]}) = Pic(K3) \oplus {\bf Z}e$.
This means that the Picard lattice of $K3^{[2]}$ can  non-trivial if
the underlying $K3$ itself is of a special type.

The Picard number ${\rho}$ of a $K3$ surface is the rank of the
Picard lattice and $K3$ surface is called attractive if ${\rho} =
20$. Defining the lattice ${\Upsilon} = (H^{2,0}(K3) \oplus
H^{0,2}(K3)) \bigcap H^2(K3;{\bf Z})$, for a generic $K3$ surface
${\Upsilon}$ is completely trivial but can have maximal rank 2. The
transendental lattice is defined as the orthocomplement of the
Picard lattice in $H^2(K3;{\bf Z})$ and precisely when the rank of
${\Upsilon}$ is two,  ${\Upsilon}$ coincides with the transcendental
lattice and the K3 surface will be attractive.

One can define an analogous lattice ${\Upsilon}$ for
hyper-K{\"a}hler four-folds also. From our discussion of ${\cal
N}=2$ supersymmetry preserving fluxes on $K3^{[2]}$ we see that
${\a}$ and ${\b}$ can be expanded in an integral basis $e_k$ for
$H^2(K3^{[2]};{\bf Z})$ as:
\eqn\sii{ {\a} = \sum_k a_ke_k \quad {\b} = \sum_l b_l e_l \quad
{\g} = {\sum}_jc_je_j \quad j,k,l = 1,...,23 }
Furthermore we can write ${G\over {\pi}} = {\sum}_{kl} N_{kl} e_k
\otimes e_l$ where we have the relation:
\eqn\siii{ N_{kl} = 2a_kb_l + 2a_lb_k - c_kc_l}

Here $a_k$ and $b_l$ are apriori real numbers and $N_{kl}$ are
integers (due to the flux quantization condition on $G$). Fixing $l$
in {\siii}, the equation {\siii} implies that a real linear
combination of ${\a}$, ${\b}$ and ${\g}$ lies on a lattice point of
$H^2(K3^{[2]};{\bf Z})$. Varying $l$ we get 23 such possibilities.
As ${\a}$, ${\b}$ and ${\g}$  are linearly independent this implies
the rank of ${\Upsilon}$ is precisely 2. That means the
corresponding $K3^{[2]}$ is what we would call attractive by analogy
with the $K3$ case.

Therefore, just as for $K3\times K3$ turning on ${\cal N}=2$
supersymmetry preserving $G$-flux leads to points in moduli space
where the underlying $K3^{[2]}$ is attractive. In the case of
$K3^{[2]}$ it is natural to associate the attractive $K3^{[2]}$ with
the attractive $K3$ from which $K3^{[2]}$ is obtained.

In fact, turning on $G$-flux of the form $G_0$ restricted the
complex structure and K{\"a}hler moduli of the hyper-K{\"a}hler
four-fold such that the complex structure was entirely fixed, and
the K{\"a}hler structure was essentially fixed upto scaling. There
is a canonical way \ref\Dijk{
  R.~Dijkgraaf,
  ``Instanton strings and hyperKaehler geometry,''
  Nucl.\ Phys.\ B {\bf 543}, 545 (1999)
  [arXiv:hep-th/9810210].} to associate an attractive $K3$ surface $S$ to a hyper-K{\"a}hler
four-fold of the form $K3^{[2]} \simeq X$. Pick a Mukai vector $v$
in $H^*(S;{\bf Z})$ with $v^2 > 0$ in the lattice ${\Lambda}^{3,19}$
of the $K3$ surface. Then the Mukai map restricted to $v^{\perp}$
gives an isomorphism between lattices of $v^{\perp}$ and $H^2(X;{\bf
Z})$ where $X$ is the moduli space of semi-stable torsion free
sheaves on $K3$ surfaces. For $v^2 = 2$, $X$ coincides with the
Hilbert scheme $K3^{[2]}$ upto a HK deformation. The positive
2-plane $P$ in $H^2(X;{\bf R})$ spanned by the real and imaginary
parts of the holomorphic 2-form ${\o}$ on $X$ is identified with the
positive 2-plane of $K3$, so the complex structure of $X$ is
determined entirely by the complex structure of the corresponding
attractive $K3$. Including the periods of the $B$-field on $K3$
determines the K{\"a}hler structure of $X$ also, as shown in
{\Dijk}.

For this reason it is not surprising that turning on ${\cal N}=2$
supersymmetric fluxes lead to a ${\Upsilon}$ that can be associated
with an attractive $K3$ surface.

\subsec{Beauville's example {\Beau}}

If $X$ is  complex torus of dimension two, then the generalized
Kummer variety $K^n(X)$ is an irreducible, holomorphic symplectic
manifold of complex dimension $2n$. For $n=2$ the corresponding
hyper-K{\"a}hler 4-fold has $b_2 = 7$ and we expect a moduli space
of the form (ignoring the moduli associated with the $h^{2,1}$
moduli):
\eqn\cyi{{\cal M}_V = {{O(4,5)}\over {O(4)\times O(5)}} }
in the absence of fluxes. As $h^{2,1}=2$ the Euler number can be
computed to be ${{\chi}\over {24}} = {{9}\over 2}$. Thus we need to
turn on flux even for these backgrounds. Let us look for solutions
with ${\cal N}=3$ supersymmetry.

We need to solve:
\eqn\cyii{ n+ {{\nu}^2\over 8}\int_X({\a}\wedge{\b})^2 = {{9}\over
2} \quad {\nu} = 2{\bf Z}+1 }

For $K^2(T^4)$ the normalization is such that:
\eqn\cyiii{ \int_X {\a}^4 = 9{q_X}^2({\a},{\a}) \quad
\int_X({\a}\wedge {\b})^2 = 3(2{q_X}^2({\a},{\b}) +
q_X({\a})q_X({\b})  ) }

This implies:
\eqn\cyiv{ \int_X({\a}\wedge{\b})^2 = 12p \quad p \in 2{\bf Z}+1 }

and:
\eqn\cyv{ n + {3\over 2}{\nu}^2p = {{9}\over 2} }

This has a solution ${\nu} = {\pm}1$ and $n=0$, where
$q_X({\a},{\b}) = 0$ and $q_X({\a}) = -6, q_X({\b}) = -2$.

As in the case of Hilb$^2(K3)$ we need to impose the condition that
$G$-flux is primitive in order for solutions to preserve
supersymmetry.

Picking a $G$-flux in $sym^2(H^2)$ to be of the form {\cuiii} we can
solve the anomaly cancelation condition as in the previous section.
Again we find ${\cal N}=3$ supersymmetric solutions with a moduli
space of the form:
\eqn\cyva{ {\cal M}_V ={{O(4,3)}\over {O(4)\times O(3)} } }

However, unlike Hilb$^2(K3)$ it is no longer true for $K^2(T^4)$
that all of $H^4$ is generated from $sym^2(H^2)$. This allows us to
consider apriori a $G$-flux of the form ${G\over {\pi}} = x$  with
$x \in (H^4(X;{\bf R}) \backslash sym^2(H^2(X;{\bf R}))\bigcap
H^{2,2}(X;{\bf Z})$.

With such a $G$-flux we need to solve:
\eqn\cyvi{ n + {1\over 8}\int_X(x^2) = {{9}\over 2} }

In order for there to be solutions to {\cyvi} we need $\int_X x^2$
to be divisible by 4. As long as $x {\in} sym^2 H^2$ this
divisibility is guaranteed by the fact that the Beauville-Bogomolov
form is even, for $K^2(T^4)$.

Suppose there is such a $x$ which solves the anomaly cancelation
condition. Then turning on a $G$-flux proportional to $x$ does not
lift any vector multiplet moduli. Indeed the simplest way to see
this is of course that the moduli space ${\cal M}_V$ was associated
to $H^2(X)$ and the $G$-flux does not depend on $H^2$. Another way
to see it is to note that given a $x$ in the ortho-complement of
$sym^2(H^2)$ in $H^4$, it is automatically primitive. However, if
$x$ belongs in the ortho-complement to $sym^2(H^2)$ (the inner
product being given by the interesection form on $H^4$) then the
interesction numbers ${\l}_{ab}$ vanish. This means turning on such
a $G$-flux does not lift any moduli and leads to a solution with
${\cal N}=3$ supersymmetry and a moduli space identical to {\cyi}
except for possible membranes on ${\bf R}^3$. Furthermore, it is not
possible to lift any hyper-multiplet moduli by turning on such a
flux.

The only question to address is whether such $x$ satisfies the
anomaly cancelation condition. For $K^2(T^4)$, $c_2$ belongs to
$sym^2(H^2)$ so that ${\l}.x = 0$. Now Wu's formula tells us $x^2 =
{\l}.x$ mod 2 so that $x^2$ is even.

We do not know how to evaluate if $x^2$ can be divisible by 4. It is
possible that there is such an $x$ for $K^2(T^4)$ in which case we
would have another ${\cal N}=3$ supersymmetric solution. The
somewhat surprising thing here would be that the flux so turned on
preserves ${\cal N}=3$ supersymmetry but does not lift any moduli at
all. We believe this unlikely, so we suspect such a flux $x$ cannot
be of type (2,2).

In order to determine solutions with ${\cal N}=2$ supersymmetry we
follow the discussion in the previous section and turn on $G$-flux
of the form {\qii}. The anomaly cancelation condition now becomes:
\eqn\cyvii{ n + {5\over 8}q^2_X({\o},{\bar{\o}}) = {{9}\over 2}}

and flux quantization requires ${\o} \in H^2(X;{\bf Z}) \bigcap
H^{2,0}(X;{\bf C})$ and ${\sqrt{{{q_X({\o},{\bar{\o}})}\over
{3q_X(J,J)}}}}J \in H^{1,1}(X;{\bf Z})$.

There is a solution to {\cyvii} which requires $n = 2$ and
$q_X({\o},{\bar{\o}}) = 2$.

\newsec{Discussion}

In this note, we have discussed the moduli space of hyper-K{\"a}hler
four-fold compactifications in type IIA/B and ${\cal M}$-theory. As
we have seen, it is possible to put a strict upper bound on the
dimension of the moduli space of hyper-K{\"a}hler four-fold
compactifications. It should not be too surprising that the moduli
spaces coming out of string theory are bounded: after all they are
related to the moduli space of the worldsheet SCFT and we have every
reason to expect this moduli space to be finite dimensional. What is
perhaps a little surprising is that one has a strict upper bound as
a mathematical result. It would be very interesting to obtain this
result from string theory arguments, or find a more intuitive reason
for this bound.

One is free to think that hyper-K{\"a}hler four-fold
compactifications only parameterize a small part of the space of all
${\cal N}=3$ vacua arising from string theory so these bounds cannot
be taken to imply anything about the bound on the moduli space of
${\cal N}=3$ vacua in string theory. We try to construct other
theories with six supercharges from string theory in appendix 2, and
show that if anything the space of ${\cal N}=3$ supersymmetric
string vacua is very tightly constrained. There are no known large
class of compactifications that achieve ${\cal N}=3$ supersymetry
with or without fluxes. The possible exception to this may be the
compactifications discussed in \ref\TsimpisKJ{
  D.~Tsimpis,
  ``M-theory on eight-manifolds revisited: N = 1 supersymmetry and generalized
  JHEP {\bf 0604}, 027 (2006)
  [arXiv:hep-th/0511047].}.

As emphasised in \ref\VafaUI{
  C.~Vafa,
  ``The string landscape and the swampland,''
  arXiv:hep-th/0509212.} the fact that the scalar moduli spaces
  arising in string theory are typically finite provides us a clue
  that consistent coupling of matter to gravity is a very nontrivial
  problem. From whatever we know about low energy supergravity there
  appears to be no reason for these moduli spaces to be bounded.

More work is needed to determine if there are some solutions of
orbifold/ orientifold type which do yield ${\cal N}=3$ supersymmetry
upon turning on fluxes. If such solutions exist, they will be
expected to belong to the moduli space of hyper-K{\"a}hler
compactifications and may perhaps allow us to understand this moduli
space better.

We have also analysed the two known examples of compact
hyper-K{\"a}hler four-folds and found that in the case of
Hlb$^2(K3)$ there is an interesting subtlety associated with the
orbifold limit $S^2({\bf T}^4/{\bf Z}_2)$.

\vskip 1cm

\newsec{Acknowledgements}
First and foremost, I would like to thank Sergei Gukov for kindly
reviewing the draft and for help with questions. I would like to
acknowledge William Linch for discussions on hyper-K{\"a}hler
manifolds, Markus Luty for his constant encouragement and help
during the completion of this work, Brenno Vallilo for kindly going
through an earlier draft and providing valuable feedback and
Sylvester James Gates Jr for discussions on ${\cal N}=(0,6)$ and
${\cal N}=(3,3)$ supergravity. I would also like to thank Dmitrios
Tsimpis for a clarifying email.

\vskip 1cm

\noindent{\bf {Appendix 1: Worldsheet aspects of hyper-K{\"a}hler
four-folds}}

The worldsheet description of ${\cal N}=(3,3)$ and ${\cal N}=(0,6)$
supersymmetric string compactifications to two dimensions starts
with an internal ${\cal N}=4$ SCFT with $c=12$. This fact is proven
as follows: In the RNS formalism supersymmetries that are gauged in
space-time must arise from global currents on the worldsheet. This
in particular means there must exist global space-time fermionic
currents of the following form:
\eqn\xi{{Q_+}^A = \int dz e^{-{\varphi}/2} e^{iH/2} {\S}^A \quad A =
1,2,3 }

{\xi} is the standard FMS vertex which is the holomorphic part of
the gravitino vertex operator at zero momentum. As usual the free
fermions ${\psi}^0$ and ${\psi}^1$ corresponding to the flat two
dimensional space-time have been bosonized into a chiral boson $H$
and {\xi} is in the standard $(-{1\over 2})$ picture so ${\S}^A$
must have dimension ${1\over 2}$. Furthermore the space-time
supersymmetry algebra without central charges is of the form:
\eqn\xii{ \{Q_+^A, Q_+^B \} = {\d}^{AB}P_+ }
which requires ${\S}^A$ to satisfy the following OPEs:
\eqn\xiii{ {\S}^A(z) {\S}^B(w)  = {\d}^{AB} {1\over {(z-w)} } }

which automatically identify ${\S}^A$ as free Majorana fermions. One
can choose a pair of fermions out of the three free Majorana
fermions and bosonize the pair as:
\eqn\xiv{ {1\over 2}({\S}^1 + i{\S}^2) = e^{i{\phi}} }

This defines a $U(1)$ current:
\eqn\xv{ J_3 = 2i{\partial}{\phi} }
which is actually an R-current. Using the remaining free Majorana
fermion ${\S}^3$ we can define two more $U(1)$ generators:
\eqn\xvi{J^{\pm} = :e^{{\pm}i{\phi}}{\S}^3:}
$(J^{\pm},J_3)$ together generate the $SU(2)$ Kac-Moody algebra at
level $k = 2$. Therefore the internal SCFT turns out to have a small
${\cal N}=4$ SCA with $c=12$. This corresponds to the case of
hyper-K{\"a}hler four-fold compactifications\footnote{$^8$}{ There
is also the possibility of getting six supercharges starting with
two commuting ${\cal N}=4$ SCFTs each with $c=6$ for the left movers
and having a ${\cal N}=2$ SCA for the right-movers. This way the
four supercharges from the left-moving sector and two from the
right-moving sector yield in total six supercharges. This way
however we end up with ${\cal N}=(4,2)$ space-time supersymmetry
from type IIA and ${\cal N}=(0,6)$ supersymmetry from type IIB.}. In
the large radius limit the worldsheet description of a
hyper-K{\"a}hler four-fold compactification is via a ${\cal N}=4$
supersymmetric sigma model which is also conformally invariant and
leads to a SCFT with small ${\cal N}=4$ SCA and $c=12$. The moduli
of the SCFT are the ${\cal N}=4$ chiral primaries.

In order to deform the ${\cal N}=(4,4)$ worldsheet SCFT one adds
operators of the form:
\eqn\xvii{ {\d}S = \int d^2z {\cal O}(z,{\bar z}) }

To preserve conformal invariance ${\cal O}$ must be a dimension
$(1,1)$ operator. However in order to preserve ${\cal N}=(4,4)$
worldsheet supersymmetry we require more. A ${\cal N}=(4,4)$ SCA has
four left-moving (and four right-moving) supercharges which can be
denoted as $G^{\pm}$ and ${\tilde G}^{\pm}$. The ${\pm}$ indices
indicate the $U(1)$ R-charge of these operators under $J_3$. In
order to preserve ${\cal N}=(2,2)$ supersymmetry generated by
$G^{\pm}$ and $J_3$ we require the operator ${\cal O}$ to be the top
component of a chiral superfield whose bottom component is a chiral
primary operator. That is, given a ${\phi}_i$ annihilated by
$G^+_{-{1\over 2}}$ and carrying charge $+1$ and conformal dimension
${1\over 2}$ one deforms the action $S$ into:
\eqn\xviii{ S' = S + \int d^2z (t^i G^- {\phi}_i +{\bar t}^i G^+
{\bar{\phi}}_i) }
where by $G^+ {\phi}_i$ one means picking the $z^{-1}$ pole of the
$G^+$${\phi}_i$ OPE. Under what circumstances will a deformation of
the form {\xviii} respect ${\cal N}=4$ superconformal invariance?
For this the deformation {\xviii} must be a $SU(2)$ singlet. It is
obviously a singlet under $U(1)$ generated by $J$ so we need to only
check invariance under $J^{\pm}$. This requires:
\eqn\xix{ {\tilde G}^+ {\phi}_i = 0}

That is ${\phi}_i$ is a ${\cal N}=4$ primary with dimension ${1\over
2}$. This is the standard result that the CFT moduli arise from
${\cal N}=4$ primary operators with dimension ${1\over 2}$.

\vskip 3cm

\noindent {\bf{Appendix 2: Other theories with six supercharges}}

Let us explore the other possibilities that give us ${\cal N}=(3,3)$
supersymmetry in 2d. As mentioned before there appears to be one
possiblity leading to ${\cal N}=(4,2)$ space-time supersymmetry in
2d which we expect to be anomalous. What are the other
possibilities?  Equation {\xi} is not the only possibility. The
other possibility is to obtain three supercharges from left movers
as:
\eqn\yi{ {Q_+}^A = \int dz e^{-{1\over 2}{\varphi}} e^{i{H\over 2}}
{\S}^A  \quad A = 1,2}

\eqn\yii{ Q_+ = \int dz e^{-{1\over 2}{\varphi}} e^{-i{H\over 2}}
{\S} }

which would give rise to ${\cal N}=(2,1)$ supersymmetry from the
left-movers. Together with the right movers one can obtain ${\cal
N}=(3,3)$ supersymmetry. Unfortunately, this will not work for the
following reason. The OPEs following from {\yi} and {\yii} imply
that ${\S}^A$ are free and can be bosonized as in {\xiv}. However
there are non-trivial OPEs between the bosonized field ${\phi}$ and
${\S}$ of the form:
\eqn\yiii{ (e^{i{\phi}}(z) \pm e^{-i{\phi}(z)}) {\S}(w) =
(z-w)^{1\over 2} \quad {\S}(z){\S}(w) = {1\over {(z-w)}}}

In a standard fashion one can decouple ${\phi}$ from ${\S}$ by
writing:
\eqn\yiv{ {\S} = \sum_q :e^{i{q\over 2}\phi}{\S}_q: }

It is easy to see that {\yiii} cannot hold for this form of ${\S}$
since there are always singularities of the form $z^{-{q\over 2}}$
and $z^{q\over 2}$ simultaneously, rather than the $z^{1\over 2}$
singularity alone that is expected in {\yiii}. In other words there
is no way to make a supersymmetry algebra of the form {\yi} and
{\yii}. This leaves us with only very few possibilities for vacua
with ${\cal N}=3$ supersymmetry in 2d. They appear upon turning on
RR fluxes or orientifolding in compactifications with higher degree
of supersymmetry.

For example we can contemplate whether starting from ${\cal N}=4$
supersymmetric compactifications in 3d we can turning on flux that
breaks ${\cal N}=4$ to ${\cal N}=3$. Unfortunately this cannot
happen, since the only ${\cal N}=4$ supersymmetric backgrounds are
either of $K3\times K3$ type or $T^2\times CY_3$ in general. For
$K3\times K3$ it is only possible to break ${\cal N}=4$ to ${\cal
N}=2$ by turning on fluxes, while ${\chi}(T^2\times CY_3) = 0$ so no
fluxes are allowed.

It is now known that the class of compactifications studied in
{\BeckerGJ} is not the most general supersymmetric solution with
${\cal N}=2$ supersymmetry \ref\MartelliKI{
  D.~Martelli and J.~Sparks,
  ``G-structures, fluxes and calibrations in M-theory,''
  Phys.\ Rev.\ D {\bf 68}, 085014 (2003)
  [arXiv:hep-th/0306225].}
. One way to interpret the solutions in {\MartelliKI} is in terms of
calibration conditions for $M5$-branes wrapping supersymmetric
cycles. For example we can consider five-branes wrapping SLAG cycles
of a CY 4-fold. Including the effect of backreaction we expect a
${\cal N}=1$ supersymmetric vacum in 3d, which is not a spin(7)
vacuum. Rather it belongs to the class studied in {\MartelliKI}.
This way, it is clear that the only way to generate ${\cal N}=3$
supersymmetric vacua in ${\cal M}$-theory is to consider vacua with
with ${\cal N}=6$ supersymmetry in 3d, and turn on $G$-flux that
breaks ${\cal N}=6$ to ${\cal N}=3$. There is no compact manifold
which upon compactification ${\cal M}$-theory yields ${\cal N} = 6$
supersymmetry in 3d.

In fact, even {\MartelliKI} turns out to obtain a restricted class
of ${\cal M}$-theory compactifications, and a more general class of
compactifications on eight-manifolds is possible\footnote{$^9$}{ The
author is grateful to Dmitrios Tsimpis for drawing this to his
attention.}{\TsimpisKJ}. It would be interesting to understand if
  in the case of ${\cal N}=3$ supersymmetry the solutions to this
  more general class of compactifications allow us to obtain ${\cal
  N}=3$ moduli spaces in 3d, of which the hyper-K{\"a}hler four-fold
  moduli space will be a strict subset.

It therefore appears that the only other way of obtaining ${\cal
N}=3$ vacua would be to start with even more symmetric spaces like
tori and orbifolds/ orientifolds thereof and turn on supersymmetry
breaking $G$-fluxes to end up with a ${\cal N}=3$ supersymmetric
vacuum in 3d. It would be nice to exhibit such examples, as they
will be expected to provide insight into the HK moduli space.

 \vskip 1cm

\listrefs

\end